\documentclass[preprint]{aastex}
\usepackage{graphicx}

\newcommand{\Hi}{\ion{H}{1}}

\newcommand{\Ni}{\ion{N}{1}}
\newcommand{\Oi}{\ion{O}{1}}
\newcommand{\Mgi}{\ion{Mg}{1}}
\newcommand{\Ali}{\ion{Al}{1}}
\newcommand{\Sii}{\ion{Si}{1}}
\newcommand{\Si}{\ion{S}{1}}
\newcommand{\Cai}{\ion{Ca}{1}}
\newcommand{\Caii}{\ion{Ca}{2}}
\newcommand{\Tii}{\ion{Ti}{1}}
\newcommand{\Tiii}{\ion{Ti}{2}}
\newcommand{\Cri}{\ion{Cr}{1}}
\newcommand{\Fei}{\ion{Fe}{1}}
\newcommand{\Feii}{\ion{Fe}{2}}
\newcommand{\Coi}{\ion{Co}{1}}
\newcommand{\Nii}{\ion{Ni}{1}}
\newcommand{\Zri}{\ion{Zr}{1}}

\newcommand\est{\fontsize{6}{6}\selectfont \mbox{est}}\normalfont
\newcommand\rave{\fontsize{6}{6}\selectfont \mbox{RAVE}}\normalfont
\newcommand\chem{\fontsize{6}{6}\selectfont \mbox{chem}}\normalfont
\newcommand\temp{$T_{\fontsize{6}{6}\selectfont \mbox{eff}}$}\normalfont
\newcommand\logg{$\log g$}
\newcommand\met{[m/H]}
\newcommand\Trave{$T_{\fontsize{6}{6}\selectfont\mbox{eff}\normalfont}^{\rave}$}
\newcommand\Lrave{$\log g^{\rave}$}
\newcommand\Mrave{[m/H]$^{\rave}$}
\newcommand\Mchem{[m/H]$^{\chem}$}
\newcommand\Fechem{[Fe/H]$^{\chem}$}
\newcommand\Achem{[$\alpha$/Fe]$^{\chem}$}

\newcommand\Mest{[m/H]$^{\est}$}

\shorttitle{The RAVE chemical catalogue}
\shortauthors{Boeche et al.}
\begin{document}

   \title{The RAVE catalogue of stellar elemental abundances: first data release}

   \author{
	C. Boeche\altaffilmark{1}, A. Siebert\altaffilmark{2}, M. Williams\altaffilmark{1}, 
	R.S. de Jong\altaffilmark{1}, M. Steinmetz\altaffilmark{1}, J.P. Fulbright\altaffilmark{3},  
	G.R. Ruchti\altaffilmark{3}, O. Bienaym\'e\altaffilmark{2},
	J. Bland-Hawthorn\altaffilmark{4}, R. Campbell\altaffilmark{5}, K.C.
	Freeman\altaffilmark{6}, 
	B. K. Gibson\altaffilmark{7}, G. Gilmore\altaffilmark{8}, E. K.
	Grebel\altaffilmark{9}, 
	A. Helmi\altaffilmark{10},
	U. Munari\altaffilmark{11}, J.F. Navarro\altaffilmark{12}, Q. A.
	Parker\altaffilmark{13,14}, 
	W. Reid\altaffilmark{13}, G. M. Seabroke\altaffilmark{15}, A.
	Siviero\altaffilmark{1,16}, F. G. Watson\altaffilmark{14}, 
	R. F. G. Wyse\altaffilmark{3}, T. Zwitter\altaffilmark{17,18}
	}          
\altaffiltext{1}{Leibniz-Institut f\"ur Astrophysik Potsdam (AIP), An der Sternwarte 16, 
D-14482 Potsdam, Germany}
\altaffiltext{2}{Observatoire astronomique de Strasbourg, Universit\'e de
Strasbourg, CNRS, UMR 7550, 11 rue de l'universit\'e, F-67000 
Strasbourg, France}
\altaffiltext{3}{Johns Hopkins University, Department of Physics and Astronomy, 
366 Bloomberg center, 3400 North Charles Street, Baltimore, MD 21218, USA}
\altaffiltext{4}{Sydney Institute for Astronomy, School of Physics A28, University of Sydney, NSW 2006, Australia}
\altaffiltext{5}{Western Kentucky University, Bowling Green, Kentucky, USA}
\altaffiltext{6}{Research School of Astronomy and Astrophysics, Australia
National University, Cotter Road, Weston Creek, Canberra, ACT 2611, Australia}
\altaffiltext{7}{Jeremiah Horrocks Institute, University of Central Lancashire, Preston, PR1 2HE, UK}
\altaffiltext{8}{Institute of Astronomy, University of Cambridge, Madingley Road, Cambridge CB3 0HA, UK}
\altaffiltext{9}{Astronomisches Rechen-Institut, Zentrum f\"ur Astronomie der Universit\"at Heidelberg,
M\"onchhofstr. 12-14, D-69120 Heidelberg, Germany}
\altaffiltext{10}{Kapteyn Astronomical Institute, University of Groningen, P.O. Box 800, 9700 AV Groningen,
The Netherlands}
\altaffiltext{11}{INAF Osservatorio Astronomico di Padova, Via dell'Osservatorio 8, Asiago I-36012, Italy}
\altaffiltext{12}{Department of Physics and Astronomy, University of Victoria, P.O. Box 3055, Station CSC, Victoria, BC V8W 3P6, Canada}
\altaffiltext{13}{Department of Physics and Astronomy, Faculty of Sciences, Macquarie University, Sydney, NSW 2109, Australia}
\altaffiltext{14}{Australian Astronomical Observatory, P.O. Box 296, Epping, NSW 1710, Australia}
\altaffiltext{15}{Mullard Space Science Laboratory, University College London, Holmbury St Mary, RH5 6NT, UK}
\altaffiltext{16}{Dipartimento di Astronomia, Universit\'a di Padova, Vicolo dell'Osservatorio 2, I-35122 Padova, Italy}
\altaffiltext{17}{University of Ljubljana, Faculty of Mathematics and Physics, Jadranska 19, 1000 Ljubljana, Slovenia}
\altaffiltext{18}{Center of excellence SPACE-SI, A\u sker\u ceva cesta 12, 1000 Ljubljana, Slovenia}

  \begin{abstract}
    We present chemical elemental abundances for $36,561$ stars observed by the
    RAdial Velocity Experiment (RAVE), an ambitious spectroscopic
    survey of our Galaxy at Galactic latitudes
    $|$b$|>25^{\circ}$ and with magnitudes in the range
    9$<I_{\mbox{DENIS}}<$13. RAVE spectra cover the Ca-triplet region
    at 8410--8795\AA\ with resolving power R$\sim$7500.  This first data
    release of the RAVE chemical catalogue is complementary to the
    third RAVE data release of radial velocities and stellar parameters, and 
    it contains chemical
    abundances for the elements Mg, Al, Si, Ca, Ti, Fe and Ni, with a
    mean error of $\sim$0.2 dex, as judged from accuracy tests performed on
    synthetic and real spectra. Abundances are
    estimated through a dedicated processing pipeline in which the curve
    of growth of individual lines is obtained from a library of
    absorption-line equivalent widths to construct a model spectrum
    that is then matched to the observed spectrum via a
    $\chi^2$-minimization technique. We plan to extend this pipeline
    to include estimates for other elements, such as oxygen and
    sulfur, in future data releases.
   \end{abstract}


\section{Introduction}\label{intro_sec}

Stars inherit the chemical patterns of the interstellar matter from
which they were born. At the end of their lives they return their
nuclear products to the interstellar medium through stellar winds and
supernovae, enriching it with heavier elements. The elemental abundance pattern
of every generation of stars thus depends on the previous one and, in
time, a sort of ``genealogical tree" develops. In principle, this allows
for the star formation history of a galaxy to be traced using stellar elemental abundances
\citep[Freeman \& Bland-Hawthorn][]{freeman}.

At the same time, the assembly history of a galaxy also leaves traces
in the kinematics of its stars. Indeed, once born, stars behave like a
collisioness fluid that spreads through phase space, generally leaving
clear tracks of their dynamical origin. For example, disrupted open
clusters generate moving groups of stars with similar kinematics that
can still be recognized long after disruption \citep[Eggen][]{eggen},
while accretion events produce transient stellar streams that closely 
track the orbit of their progenitor satellites \citep[Helmi et
al.,][]{helmi}.

Chemical and kinematic information, when available, can therefore be
used to reconstruct the history of the Milky Way much in the way
archaeologists examine relics to recreate the ancient
past. The data requirements of this exercise, however, have limited
its applicability in the past: assembling a large and homogeneous set
of proper motions, distances, radial velocities, and elemental
abundances requires large photometric and spectroscopic surveys that
have not been feasible until recently. 

Spectroscopy has traditionally been the bottleneck, with much
of the work on chemical elemental abundances limited to small,
biased samples. Until recently, iron abundances ([Fe/H]) had been
estimated for samples as large as the $\sim 16,000$ stars of the
Geneva-Copenhagen Survey \citep[Nordstr\"om et al.,][]{nordstrom}, but
the numbers of stars with abundances measured for multiple elements
were much smaller. Indeed, the largest homogeneous sample available up
to now was published by Valenti \& Fisher \citep{valenti}, who
measured the abundances of 5 elements in $1,040$ nearby F, G and K
stars.

Smaller samples, often composed of a few hundred stars or fewer, have
also been presented by several authors (Edvardsson et al., \citealp{edvardsson},
Fuhrmann \citealp{fuhrmann1998}, \citealp{fuhrmann2008}, Luck \&
Heiter \citealp{luck2006}, \citealp{luck2007}, Reddy et al.
\citealp{reddy}, among others). Larger, but inhomogeneous, datasets
have been collated from the literature by Soubiran \& Girard
\citep{soubiran} and Venn et al. \citep{venn}, who compiled 743 and
821 stars, respectively, and also in the PASTEL catalogue
\citep[Soubiran et al.,][]{soubiran2010}, which is a collection of 865
literature studies.

The availability of dedicated telescopes and multi-object
spectrographs has radically changed this state of affairs, enabling
surveys such as the RAdial Velocity Experiment, RAVE \citep[Steinmetz
at al.][]{steinmetz}, and the Sloan Extension for Galactic
Understanding and Exploration (SEGUE) \citep[Yanny et
al.,][]{yanny}. Combined these two surveys have now taken spectra for
roughly a million stars. Galactic archaeology has thus become one of
the fastest-growing fields of astronomical enquiry, as evidenced by
the numerous surveys currently underway or in the advanced planning
stages, e.g., Gaia \citep[Perryman at al.][]{gaia}; the Large Sky Area
Multi-Object Fiber Spectroscopic Telescope
(LAMOST\footnote{http://www.lamost.org/website/en}); and
HERMES \citep[Freeman \& Bland-Hawthorn][]{freeman1}.

We present here the first release of the chemical catalogue for RAVE,
an ambitious spectroscopic survey of our Galaxy at Galactic latitudes 
$|$b$|>25^{\circ}$ and magnitudes in the range
9$<I_{\mbox{DENIS}}<$13. This catalogue contains multi-element
abundance measurements for $36,561$ stars of the Milky Way based on
$37,848$ RAVE spectra covering the Ca-triplet region at 8410--8795\AA\
with resolving power R$\sim$7500. This data release is associated to the 
RAVE third data
release \citep[DR3, Siebert et al.][]{siebert}, where further
information (kinematics, photometry, etc.) can be found.
As of summer 2011, RAVE has taken more than half a million spectra for some
400,000 stars. Abundances for these targets will be published in
subsequent data releases.

RAVE \citep[Steinmetz et al.][]{steinmetz} was first conceived as a
radial velocity survey to provide the missing third velocity component
for stars in the solar suburb.  However, it soon became 
clear that RAVE spectra carry much more information than just
radial velocities. After further development, the RAVE processing
pipeline was modified to deliver estimations of the values of the stellar
parameters like effective temperature, gravity and metallicity (\citealp[Zwitter et
al.][]{zwitter}, \citealp[Siebert et al.][]{siebert}).

We take this development one step further here by tackling the
measurement of chemical elemental abundances at the medium spectral resolution
provided by RAVE data. This feature is shared with other large
spectroscopic surveys with limited resolution and spectral coverage,
such as Gaia and LAMOST. Although high precision is not expected from
medium resolution spectroscopy, the availability of hundreds of
thousands of RAVE spectra enables the creation of a large and
homogeneous catalogue of chemical abundances suitable for statistical
investigation.

The RAVE chemical pipeline measures abundances for 7 elements: Mg, Al,
Si, Ca, Ti, Fe and Ni.  Homogeneity is assured by an automated
processing pipeline that measures abundances assuming the stellar
parameters computed by the RAVE pipeline described in Siebert et al.
\citep{siebert}.  RAVE stars have astrometry from different sources
like Tycho2 \citep[H\o g et al.][]{hog}, PPM-Extended catalogue PPMX
\citep[Roeser et al.][]{roeser} and the Second U.S. Naval Observatory
CCD Astrograph Catalog UCAC2 \citep[Zacharias et
al.][]{zacharias}. These, together with RAVE radial velocities and
distance estimates (\citealp[Breddels et al.][]{breddels},
\citealp[Zwitter et al.][]{zwitter2010}, \citealp[Burnett et
al.][]{burnett}), yield 3D positions and velocities. Combining this
information with chemical abundances results in a unique
chemo-kinematic dataset suitable for investigations of the formation
history of the Galaxy.

This paper is structured as follows. In Section~\ref{sum_pipe_sec} we
present the chemical processing pipeline, detailing how the chemical
elemental abundances are measured.  Tests to establish the accuracy and
reliability of the results are detailed in Section~\ref{valid_sec}, while in
Section~\ref{discuss_sec} we discuss our method of measurement and in
Section~\ref{sec_catalogue} we present the RAVE chemical catalog.  In
Section~\ref{concl_sec} we outline our conclusions.

\section{The Pipeline}\label{sum_pipe_sec}

The RAVE chemical abundance pipeline uses a different approach to classical elemental abundance estimation methods. 
These methods either (i) measure equivalent widths (EW)s and infer
elemental abundances from the curves-of-growth (COG)s or (ii) synthesize spectra with
varying elemental abundances to find the best match between the synthetic and
observed spectra. The first method cannot be successfully
applied to the RAVE spectra because it requires isolated lines for
precise EW measurements and, at the medium resolution of RAVE, most of
the lines are instrumentally blended.  Also, the second method is computationally too expensive as it requires the
synthesis of several spectra $\sim$400\AA\ wide, each having hundreds of
lines. 

The adopted method can be considered a hybrid approach, 
measuring the elemental abundances by fitting the spectrum with a model
constructed with lines of known EWs. The construction of the model is robust and fast, with the elemental abundances determined by a $\chi^2$
minimization routine. The drawback of this method is that the EWs of
the lines are computed by neglecting the opacity of the neighboring
lines. This may overestimate the EWs, leading in some cases to
underestimated elemental abundances. We show below that this effect is small for
most of the lines in the RAVE wavelength range.

\subsection{Procedure}\label{pipe_sec}

The chemical pipeline utilizes RAVE reduced spectra and their stellar parameters, such as effective
temperature (\Trave), gravity (\Lrave) and metallicity (\Mrave), returned
by the RAVE pipeline (described in Siebert et al.,
\citealp{siebert})\footnote{\Mrave indicates the
uncalibrated metallicity as defined in \citealp[Zwitter et al.][]{zwitter}.}. The pipeline also uses an EW library based on
the RAVE line list. The RAVE line list contains 604 known absorption lines
identified in the RAVE wavelength range (Sec.\ref{linelist_sec}). The
EW library contains the EWs of absorption lines for every point on the \temp, \logg\ and \met\ grid (Sec.~\ref{EWlibrary_sec}), 
which accounts for the variation of the EW with the stellar parameters. 
Variations of EW with elemental abundances are then further calculated for 
each point of this grid, for five different abundance levels in the range $[-0.4,0.4]$ with respect
to the value of metallicity [m/H]. 

The pipeline contains some auxiliary codes to (re)normalize the spectrum and detect spectral
defects like bad normalization, cosmic rays and other unwanted features.

The pipeline algorithm can be outlined as follows:
\begin{enumerate}
\item Upload the normalized, RV corrected and wavelength calibrated spectrum and
the estimated stellar parameters \Trave, \Lrave, \Mrave
\item Upload the RAVE line list from the EW library and the corresponding EWs for the stellar parameters at the five different abundance levels 
\item Extract  a shorter line list of those lines which, at the estimated stellar
parameters, have large enough EWs to be visible above the noise
\item Fit the strong lines and correct the continuum
\item Construct the COGs of the lines by fitting a polynomial function through the five EW-abundance points
\item Create the model by assuming a Gaussian profile for each line and summing these profiles together 
\item Vary the chemical elemental abundances to obtain different models by changing the EWs of the lines according to
their COG
\item Minimize the $\chi^2$ between the models and the observed spectrum to find the best-matching model
\end{enumerate}

In the following we give further details pertaining to these steps.

\subsection{Line list and constraining the oscillator strength}
\label{linelist_sec}


The RAVE line list contains 604 absorption lines identified in spectra
of the Sun and Arcturus. The lines correspond to the element species \Ni,
\Oi, \Mgi, \Ali, \Sii, \Si, \Cai, \Tii, \Tiii, \Cri, \Fei, \Feii, \Coi, \Nii,
\Zri\ and 
to the CN molecule. In
order to get precise chemical elemental abundances from the EWs we firstly need reliable atomic
parameters for the lines. A critical parameter 
is the oscillator strength ($gf$, often expressed as logarithm $\log
gf$). If precise laboratory measurements are missing for the $\log
gf$ values then they are obtained through an inverse spectral analysis to obtain `astrophysical $\log
gf$s'. This is the case for the RAVE wavelength region, where the VALD database
\citep[Kupka et al.][]{vald} reports that only 11 \Fei lines have
laboratory oscillator strength measurements. The Sun and Arcturus's spectra were synthesized and
the $\log gf$ values for each line obtained using an automated procedure which varies the
$\log gf$s so as to match both spectra simultaneously.
For the synthesis we adopted Solar abundances as given by Grevesse \& Sauval
\citep[][]{grevesse}. They will be the zero point of the abundances of this
work.
The details of the procedure are outlined in Appendix~\ref{app_linelist}.

The pipeline makes use of a shorter working line list which contains the
lines strong enough to be visible above the noise.  The lines deemed visible
are those satisfying: 
\begin{displaymath} 
EW(\mbox{m\AA})> 2\cdot\frac{1}{STN}\cdot d\cdot 1000 
\end{displaymath} 
where STN is the
signal-to-noise ratio\footnote{In the following we
  will use STN to indicate the signal-to-noise ratio as computed in
  Siebert et al. \citep[][]{siebert} whereas S/N will be used with
  the usual meaning given in spectrophotometry.} and $d$ is the spectral dispersion (for RAVE
spectra $d=0.4$\AA/pix). Other lines can blend with these visible lines, contributing to their flux absorption and changing their
apparent EW. We therefore added to the working line list all those lines
having EW$>$1m\AA\ and being closer than 0.6\AA\ (1.2\AA\ is the full-width half
maximum of a typical RAVE spectrum) to the visible lines.

\subsection{The EW library and COG computation}
\label{EWlibrary_sec}

In order to compute the line COGs we built an EW library
with the RAVE line list for $30,640$ atmosphere models. The EWs were computed using the LTE line-analysis
software MOOG \citep[Sneden][]{sneden} and the ATLAS9 model atmosphere grid \citep[Castelli \& Kurucz]{castelli}.  
As the spacing of the ATLAS9 grid is different to our requirements, we linearly interpolated the ATLAS9 grid to produce our own 
grid, covering the \temp\ range [3600,7600]K in steps of 100K, gravity
range [$0.0$,$5.0$] dex in steps of $0.2$ dex, and metallicities in the
range [$-2.5$,$+0.5$] dex in steps of $0.1$ dex.  We did not compute
the EWs for atmospheres with \temp$>$5100K and \logg$<$1.0 (hot
supergiants), because MOOG does not produce
results for these parameters and furthermore there are no such stars in our
catalogue. For every atmosphere model we computed the EWs of the lines
for 5 abundance levels with respect to iron: [X/Fe]= $-0.4$, $-0.2$, $0.0$,
$0.2$, $0.4$ dex assuming [Fe/H]=[m/H]. The whole EW library consists
of $145,080$ files. To obtain EWs for stellar parameters
between the grid points of the EW library we linearly interpolate the closest points on the grid.

\subsection{Microturbulence}

\begin{figure}[t]
\centering
\includegraphics[bb=97 488 268 670,width=6cm]{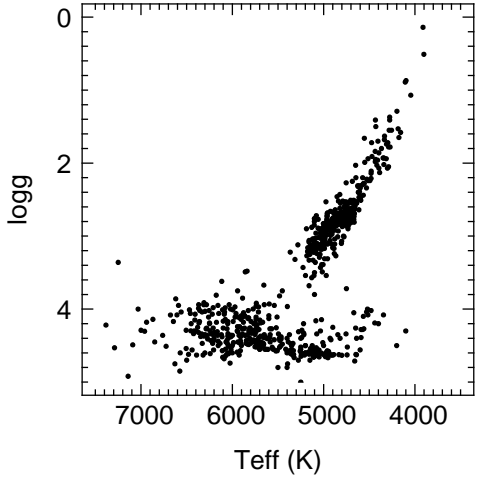}
\caption{Distribution on (\temp , \logg) plane of 712 stars observed
spectroscopically at high resolution.}
\label{microt_T_logg}
\end{figure}

The EW computation requires a value of the microturbulence $\xi$ for each atmosphere model. For high-resolution data $\xi$ is
typically determined by measuring the EWs of \Fei lines and changing $\xi$ until the iron abundances inferred from strong and weak lines
agree. As we do not measure individual EWs for our spectra we cannot use this procedure, but
instead rely on relations giving $\xi$ as a function of
the stellar parameters. Such relations have been derived by Edvardsson et al.  \citep[][]{edvardsson}, Reddy et al.,
\citep[][]{reddy2003}, Allende Prieto et al., \citep[][]{allende},
where $\xi$ is given as a function of \temp\ and \logg. Unfortunately, these results cover only specific regions of parameter
space (e.g., hot dwarfs or cold giants). Thus, we derived our own
relation covering a wide a range of \temp\ and \logg. To do so we made use of literature results from high resolution spectroscopy 
that report both stellar parameters and $\xi$ of their stellar samples. We collected data for 712 giants and
dwarfs spanning a wide range in \temp\ and \logg\  from Luck \& Heiter \citealp[][]{luck2006}, \citealp[][]{luck2007}, Bensby
et al. \citealp[][]{bensby2005}, Fuhrmann \citealp[][]{fuhrmann1998},
Fulbright et al.  \citealp[][]{fulbright2006}, Allende Prieto et
al. \citealp[][]{allende}. Figure~\ref{microt_T_logg} displays the coverage in the \temp\ - \logg\ plane for
the sample.  A 3rd-degree polynomial fit was used to obtain the
microturbulence dependence on gravity and effective temperature.  Appendix
\ref{app_microt} gives the coefficients of this polynomial fit,
$\xi_{poly}$.

\begin{figure}[b]
\centering
\includegraphics[width=12cm]{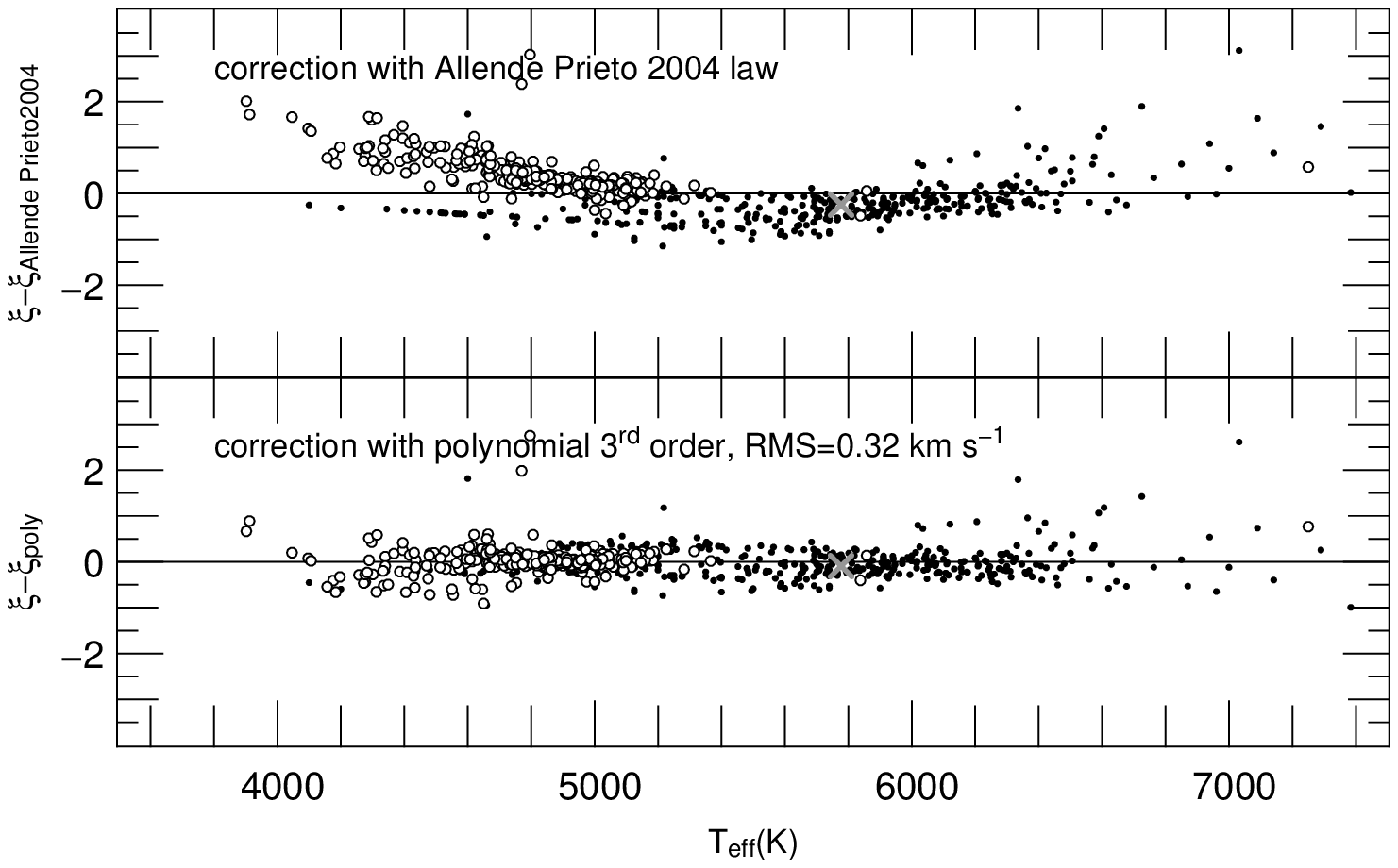}
\caption{Top panel: residuals between measured microturbulence 
$\xi$ and computed $\xi$ with
the Allende Prieto et al.'s formula for 712 stars. Bottom panel: residuals between
measured $\xi$ and computed by a 3rd degree polynomial function. Solid and
open points represent dwarf and giant stars, respectively. The gray cross
represents the Sun.
}
\label{microt_correction}
\end{figure}

Figure~\ref{microt_correction} compares the Allende Prieto et
al. \citep[][]{allende} law with $\xi_{poly}$.
Uncertainties of $\sigma_{\xi}$=0.32 km s$^{-1}$ in our polynomial law
translate approximately into $\sim 0.04$ dex elemental abundance uncertainties
for dwarfs stars (as estimated by Reddy et al. \citealp[][]{reddy2003}
and Mishenina et al. \citealp[][]{mishenina}).

\subsection{Continuum re-normalization and strong line fitting}
\label{cont_sec}

Before measuring EWs, the spectrum's continuum needs to be re-normalized and
its strong lines fitted.  For while the spectrum is continuum normalized
by the RAVE pipeline, we have found that this normalization is is not
rigorous enough for elemental abundance estimation: the RAVE pipeline
employs a 3rd order spline function to fit the continuum leaving behind
fringing effects on scales of less than 50\AA\ wide seen in some
RAVE-normalized spectra.  The chemical pipeline performs a new normalization
to remove such unwanted features, fitting the continuum and the strong \Caii
and \Hi Paschen lines.  These strong lines are then excluded from the
measurement process because they are difficult to fit properly.  The
pipeline then considers as ``continuum'' the fit to the classical continuum
plus the strong lines, and by comparison with this level the metallic lines
are measured.

Before applying the continuum correction we estimate a preliminary
metallicity, which we call \Mest\ to distinguish it from the final
metallicity \Mchem.  This preliminary metallicity is required because the
intensity of the absorption lines must be known in order to subtract them. 
When they are subtracted the continuum level can then be seen.  However, in
order to measure the intensity of a line one must know where the continuum
lies.  The continuum level is therefore determined iteratively, starting
with the normalization of the RAVE pipeline and measuring the line intensities
using the chemical elemental abundances estimation subroutine (see
Sec.~\ref{new_pipe_abd_est}), with the difference that all the abundances
vary together as one variable [X/H]=\Mest.  Wavelength intervals
$\sim$20\AA\ wide centered on the strong lines are avoided as the large
wings of these lines can affect the results of the $\chi^2$ determination of
the metallicity.  The best-match \Mest\ is then used to synthesize the
metallic lines, which are then subtracted before the strong lines are fit.

\begin{figure}[t]
\centering
\includegraphics[width=12cm]{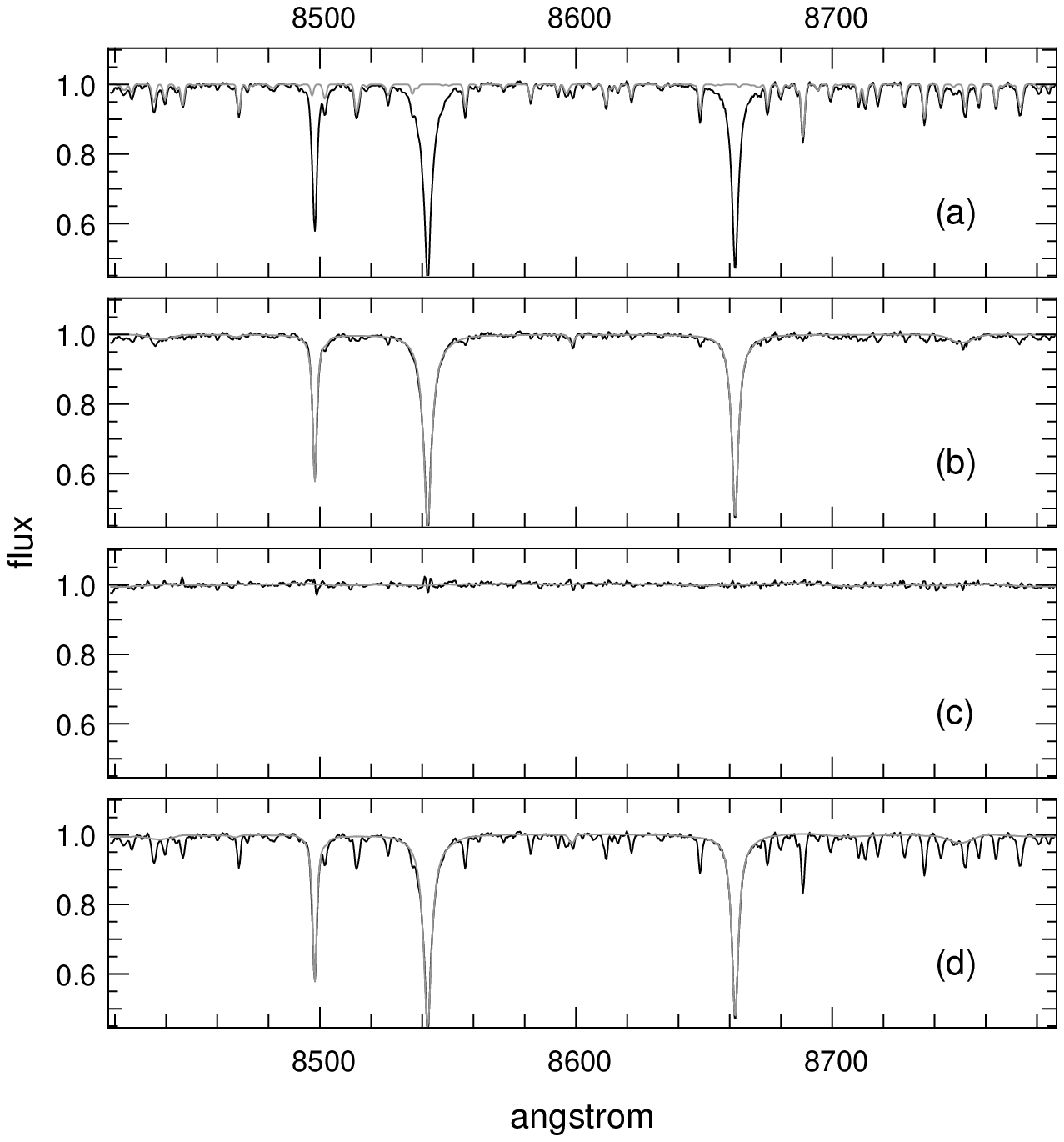}
\caption{Black line: RAVE spectrum of the Sun. Panel (a): metallic lines
fitting. Panel (b): strong line fitting. Panel (c): continuum fitting. 
Panel (d): the grey line is the sum of the gray lines in panels (b) and (c). 
This curve is used as the continuum line from which we estimate the chemical elemental abundances.}
\label{cont_sun}
\end{figure}

Fitting the continuum is performed in four steps, summarized below and illustrated in
Figure~\ref{cont_sun} from top to bottom:
\begin{enumerate}
\item Perform a preliminary metallicity and subtract the metallic lines (gray line in panel (a), 
  Figure~\ref{cont_sun}).
\item Fit the strong lines with a Lorentzian profile for \Caii and
  a Gaussian profile for \Hi and subtract them (gray line in panel (b), 
  Figure~\ref{cont_sun}).
\item Estimate the continuum profile by box-car smoothing what remains (gray line in panel (c),
Figure~\ref{cont_sun}).
\item Add the strong lines and the continuum together to obtain the new ``continuum"
  (grey line in panel (d), Figure~\ref{cont_sun})
\end{enumerate}

\subsection{Chemical elemental abundance estimation}
\label{new_pipe_abd_est}

The chemical pipeline uses the parameters \Trave\ and \Lrave\ for the
elemental abundance estimation.  The observed spectrum is fit by a
model spectrum firstly obtained
by subtracting the flux absorbed by the metallic lines to that derived above
in Sec.  \ref{cont_sec}, i.e., the strong lines + continuum fit (gray line
in panel (d), Figure~\ref{cont_sun}).  The absorption lines are assumed to
have Gaussian profiles, and as the instrumental profile is dominant with
respect to the line profile, we use the same full width half maximum (FWHM)
for all lines.  However, this FWHM is varied and optimized during the
$\chi^2$ minimization process.

The pipeline computes the COG of the lines by using the EWs at five
levels of abundances ([X/Fe]= $-0.4$, $-0.2$, $0.0$, $0.2$, $0.4$ dex,
assuming [Fe/H]=\Mrave) from the EW library. For every line,
the five EW points are fit with a 3rd order polynomial function,
which serves as the COG. However, these polynomial functions represent only 
part of the COGs and will diverge if extrapolated too far beyond the five levels in
the EW library. To avoid such divergences we limit accepted abundance results to those in the range
$-0.6 \leq$[X/Fe]$\leq +0.6$\footnote{Abundances beyond this limit can be
due to photon noise stronger than the absorption lines or by peculiar
abundances. The two cases cannot be distinguished by the chemical pipeline.}. 

Using the COGs the pipeline then creates the model spectrum, using 
[X/H]=\Mrave\ as a first-guess to the metallicity. It computes the
$\chi^2$ between the observed spectrum and the model and, through a
minimization process, changes the elemental abundances [X/H] until the best
match (minimum $\chi^2$) with the observed spectrum is reached. The
minimization process is performed with 15 variables: 13 elemental abundances,
one molecule and the instrumental FWHM of the lines.

\begin{figure}[t]
\centering
\includegraphics[width=12cm]{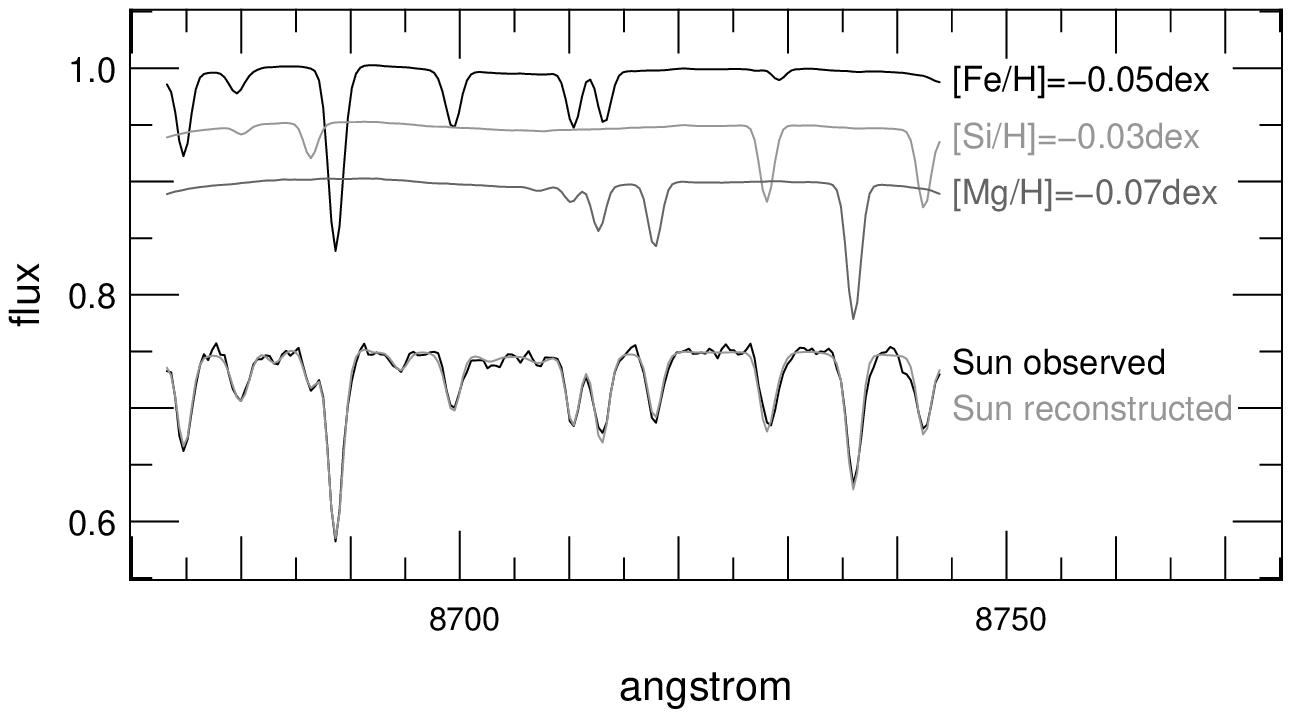}
\caption{Once the abundances of Fe, Si and Mg are fixed their absorption lines can be
synthesized (gray lines on the top). By summing them
together we obtain the grey line on the bottom, which is the model spectrum of the
Sun. The black line is the Sun's spectrum reflected by the Moon and observed by RAVE.}
\label{reconstr_sp}
\end{figure}

Figure~\ref{reconstr_sp} shows the best-match model
spectrum of the Sun and compares it with the observed one. The three spectra on the
top show how the model spectrum is built up: the spectra of
three elements (Fe, Si, Mg) at parameters \temp\ =5861K and \logg\
=4.54 are generated according to the estimated elemental abundances and
added together to construct the solar spectrum.

This methodology has a drawback:
the computed COG of any absoption line neglects the 
opacity of the neighboring lines, leading to an overestimation of the EW
of blended lines and so an underestimation of the elemental abundances. 
This systematic originates from the EW library created by using the MOOG's driver
``ewfind", which computes the expected EW of the lines as if they were isolated.
In order to minimize this systematic error we apply a
correction coefficient to reduce the EWs of lines which are physically
blended. The corrected EW
is given by
\begin{equation}\label{opacity_for}
EW_{\mbox{corr}}=EW \cdot coeff \cdot cont,
\end{equation} 
where the coefficient
\begin{displaymath}
coeff=1-\sum_i^{neighbor<0.2\AA} EW_i/2.50/dispersion,
\end{displaymath} 
is computed by considering all the neighbouring lines within 0.2\AA\ 
of the line of interest. The dispersion is expressed in \AA /pix and $cont$ is
the value of the continuum (gray line in panel (d) of
Figure~\ref{cont_sun}) at the central wavelength of the line.  The
multiplication of the EW by $cont$ in Eq.~\ref{opacity_for}
corrects for the effects of strong
lines (such as \Caii), if the line of interest lies within their large wings.  This
correction reduces the systematic error in the abundances from $\sim -0.15$ dex to $\sim
-0.1$ dex or less for most elements (see details in the following
sections, where quality checks are outlined).

\subsection{Spectrum quality determination: the {\it frac} parameter}
\label{mask_sec}

Roughly 25\% of RAVE spectra are affected by defects like fringing or
cosmic rays that cannot be removed by continuum correction (they usually
affect regions smaller than 100\AA).  In order to determine the locus and size of
the defect we define the flux
residual between the observed and the best-match model for the $i$-th
pixel with
\begin{displaymath}
r(i)=f_{model}(i)-f_{obs}(i),
\end{displaymath}
where $f$ is the flux of the spectra, and we
use the following algorithm (also used in Siebert et al.,
\citealp{siebert}):
\begin{enumerate}
\item Consider the interval $I_j=[j-10,j+10]$ centered on the pixel
  $j$. Compute
\begin{displaymath}
\tilde{\chi}^2(j) =\frac{1}{max(I_j)-min(I_j)}\cdot
 \sum_{i=min(I_j)}^{max(I_j)}(\frac{r(i)}{\sigma})^2,
\end{displaymath}
\begin{displaymath}
\psi(j) =\frac{1}{max(I_j)-min(I_j)}\sum_{i=min(I_j)}^{max(I_j)} r(i),
\end{displaymath}
where $\tilde{\chi}^2(j)$ is the reduced $\chi^2$ and $\psi(j)$ is the
estimation of the area between the observed and model
spectra. $\sigma$ is the inverse of the signal-to-noise (STN) ratio.
\item If $\tilde{\chi}^2(j)>2$ and $\psi(j)>2\cdot \sigma$ then the
  pixels in the interval $I_J$ are labeled as a defect.
\end{enumerate}

The process is repeated for all the pixels of the spectrum, resulting
in an array of 0 and 1, with 1 label indicating non-defective
pixels. The fraction of pixels that are non-defective is then denoted 
with the {\it frac} parameter (i.e., the higher the {\it frac}
parameter, the better the quality of the spectrum). We deem spectra with $frac<0.7$ 
as overly affected by fringing/cosmic ray defects and exclude these spectra from our analysis.

\newpage
\section{Validation and accuracy}\label{valid_sec}

There are several sources of error that affect the elemental abundances
measurement process. For example, there are uncertainties in (i) the oscillator
strength values, (ii) the continuum normalization, (iii)
the stellar parameters; (iv) uncertainties in
microturbulence, (v) uncertainties due to local line opacity neglected
by the pipeline, (vi) neglect of non-LTE effects and (vii) photon
shot noise. These sources of errors interact in a complex fashion,
making the error estimation process quite challenging. To establish
the capability of the pipeline to derive the chemical elemental abundances we
ran a series of tests on synthetic and real spectra for which stellar
parameters and elemental abundances are well known. The results are presented
below.

\subsection{Chemical elemental abundances accuracy from synthetic
spectra}\label{test_synt}

\begin{figure}[t]
\centering
\includegraphics[width=6cm]{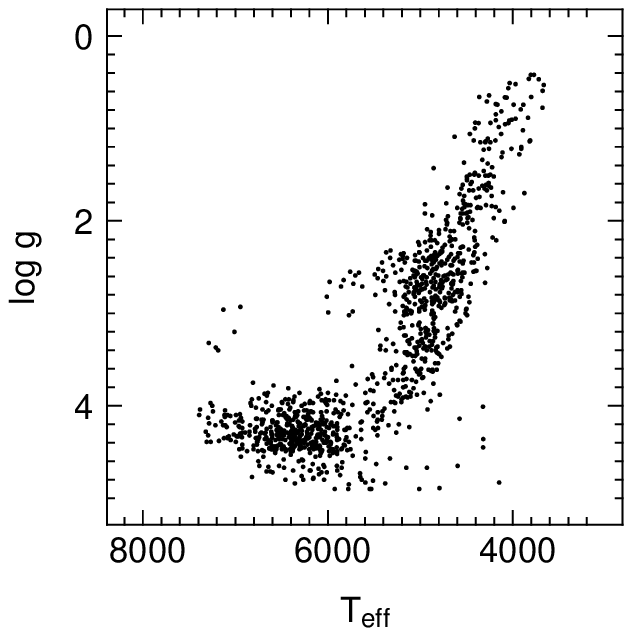}
\caption{Distribution of the synthetic spectra sample on the stellar
parameters plane \temp\ and \logg.}
\label{besancon_CMD}
\end{figure}

We ran the pipeline on a sample of $1,353$ synthetic spectra to
which three intensity levels of artificial noise were added, testing 
the accuracy of the results at S/N$=100$, $40$, and $20$. To make the
sample as realistic as possible, the spectra have been synthesized with
distributions of \temp\ and \logg\ from a mock sample of RAVE
observations (M.  Williams, private communication) created by using the
Besan\c con model \citep[Robin et al.][]{robin}. The \temp\ vs \logg\
distribution of the sample is shown in Figure~\ref{besancon_CMD}.

Each spectrum of the sample had the elemental abundances of one of the
stars from the Soubiran \& Girard catalogue \citep[][]{soubiran}, 
ensuring realistic distribution of the chemical abundances.  As
our line list has only eight elements in common with the Soubiran \&
Girard catalogue, we assigned the elemental abundance [X/H]=[Fe/H] to the
elements Cr, Co, Ni, Zr. We also assigned [X/H]=[$\alpha$/H] to C, N, O, Mg, Al, Si,
S, Ca, Ti when the measurement of one of these elements was
missing. The spectra were synthesized using the code MOOG at
resolution 0.01\AA/pix and degraded to RAVE resolution (0.4\AA/pix,
1.4\AA\ FWHM).

The first test was performed by adopting the stellar parameters \temp,
\logg\ and \met\ of the synthetic spectrum in order to evaluate the
accuracy of the elemental abundance measurements of the pipeline alone. The
results are illustrated in Figure~\ref{abd_comparison}.  For the second
test we added uncertainties to the stellar parameters so as to simulate real RAVE spectra.  Following the typical RAVE
errors, we added a random Gaussian error with standard deviation 300K
in \temp , 0.5 dex in \logg\ and 0.3 dex in \met\ to the correct
parameters and ran the pipeline adopting such ``wrong" parameters.
For the sake of brevity and clarity we present only the results
for the seven most reliable elements (see also
Figure~\ref{abd_comparison_err}) .

\subsubsection{Results assuming no error in the stellar parameters} 

{\bf Results at S/N=100.}  With no errors added to the stellar
parameters the residuals between measured and expected elemental abundances [X/H] 
have a systematic error of $\simeq -0.10$ dex and standard deviation 
$\sigma \simeq 0.10$ dex or smaller
(see Figure~\ref{abd_comparison}), with a slight variation from
element to element. The relative enhancement [X/Fe] shows a small 
($\simeq -0.10$ dex) systematic error at high [X/Fe].

{\bf Results at S/N=40.}  The elemental abundance residuals have 
$\sigma \simeq$0.10-0.20 dex depending on the element and show the same systematics seen with
S/N=100, but slightly less pronounced.

{\bf Results at S/N=20.}  At this S/N the pipeline can still estimate
abundances of Fe, Al and Si and the relative enhancement to iron
[X/Fe], but with $\sigma \simeq$0.2-0.3 dex. Other
elements like Mg and Ti exhibit larger systematic errors.  An
$\alpha$-element abundance estimated as the average of Mg and Si still
yields reliable results with an error of $\sigma\simeq 0.2$ dex.
Elements such as Ca and Ni cannot be reliably measured because the
lines are too weak to be detected.

\begin{figure}[t]
\centering
\includegraphics[width=12cm]{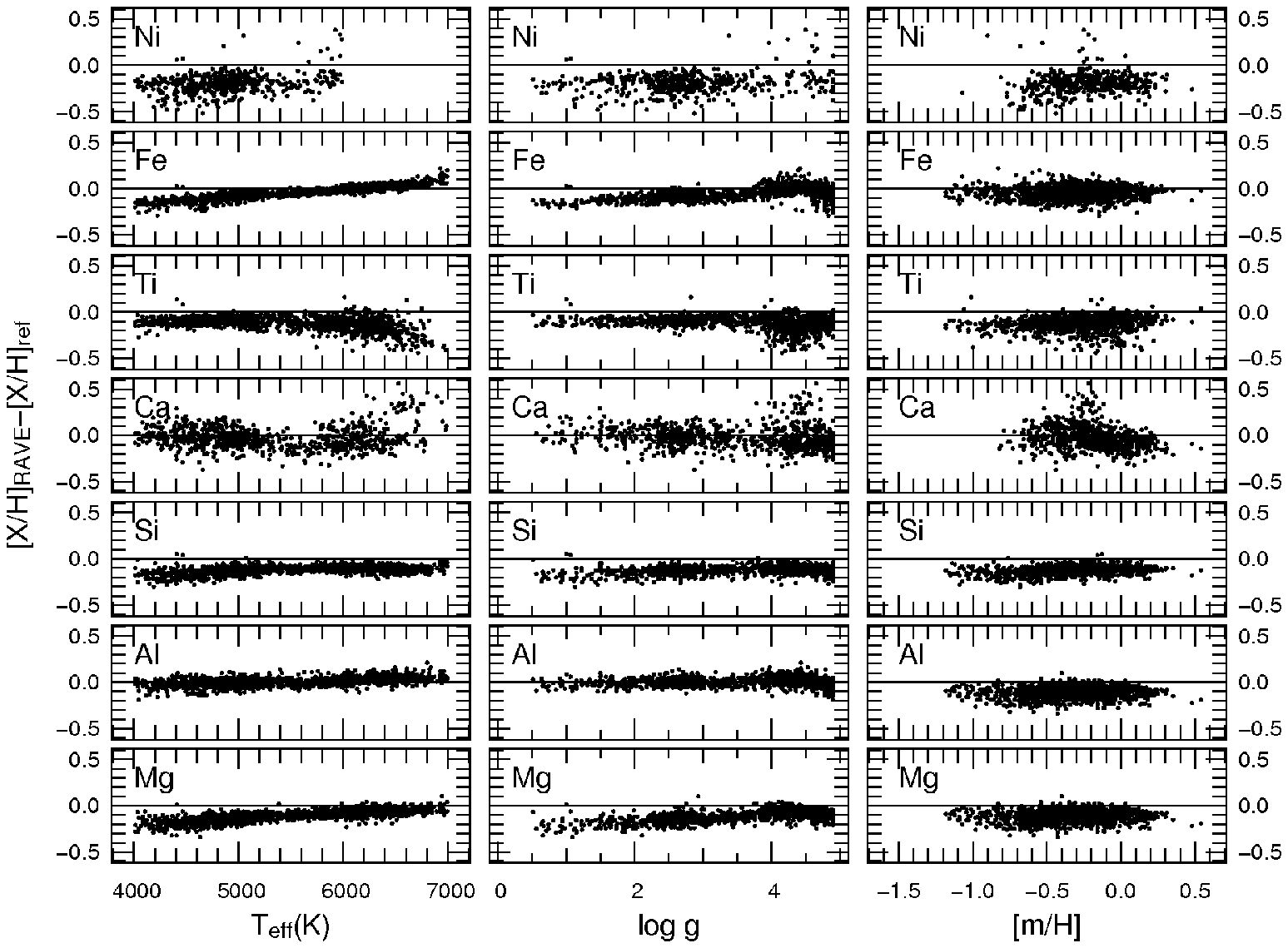}
\caption{Correlation between the elemental abundance residuals measured-minus-expected
(y-axis) and the stellar parameters (x-axis) at S/N=100.}
\label{resid_abd_params}
\end{figure}

We found correlations between stellar parameters and elemental chemical
abundances obtained. In particular, the abundances correlate with \temp\
(Figure~\ref{resid_abd_params}). This is very likely due to continuum
correction, which can appear lower than the real level in spectra crowded of
lines, as in cold giants stars.

\subsubsection{Results including errors in the stellar parameters}

{\bf Results at S/N=100.}  When errors are added to the stellar
parameters the elemental abundances [X/H] are affected to the level of $\sigma \simeq
0.15-0.20$ dex for elements like Mg, Al, Si, Ca, Fe, Ni.  Titanium
abundances show clear systematic errors, being overestimated at high
abundance and underestimated at low abundance. Enhancements relative
to Fe, [X/Fe], have errors $\sigma<0.2$ dex and little sign of
systematic errors for most of the elements.

{\bf Results at S/N=40.}  The elemental abundance errors are in the range
$\sigma \simeq$0.15-0.30 dex depending on the element. The relative
elemental abundances [X/Fe] are reliable with errors $\sigma \simeq$0.2 dex.

{\bf Results at S/N=20.} 
At this S/N the pipeline can still estimate abundances of
Fe, Al and Si and the relative enhancement with respect to iron [X/Fe]
even with an error of $\sigma \simeq$0.2-0.3 dex.
Other elements like Mg and Ti show systematic errors.
An $\alpha$-element abundance computed by averaging Mg and Si yields
reliable results with an error of $\sigma\simeq 0.2$ dex.
Elements such as Ca and Ni cannot be reliably measured because the lines
are too weak to be detected.

\begin{figure}[t]
\centering
\includegraphics[width=12cm]{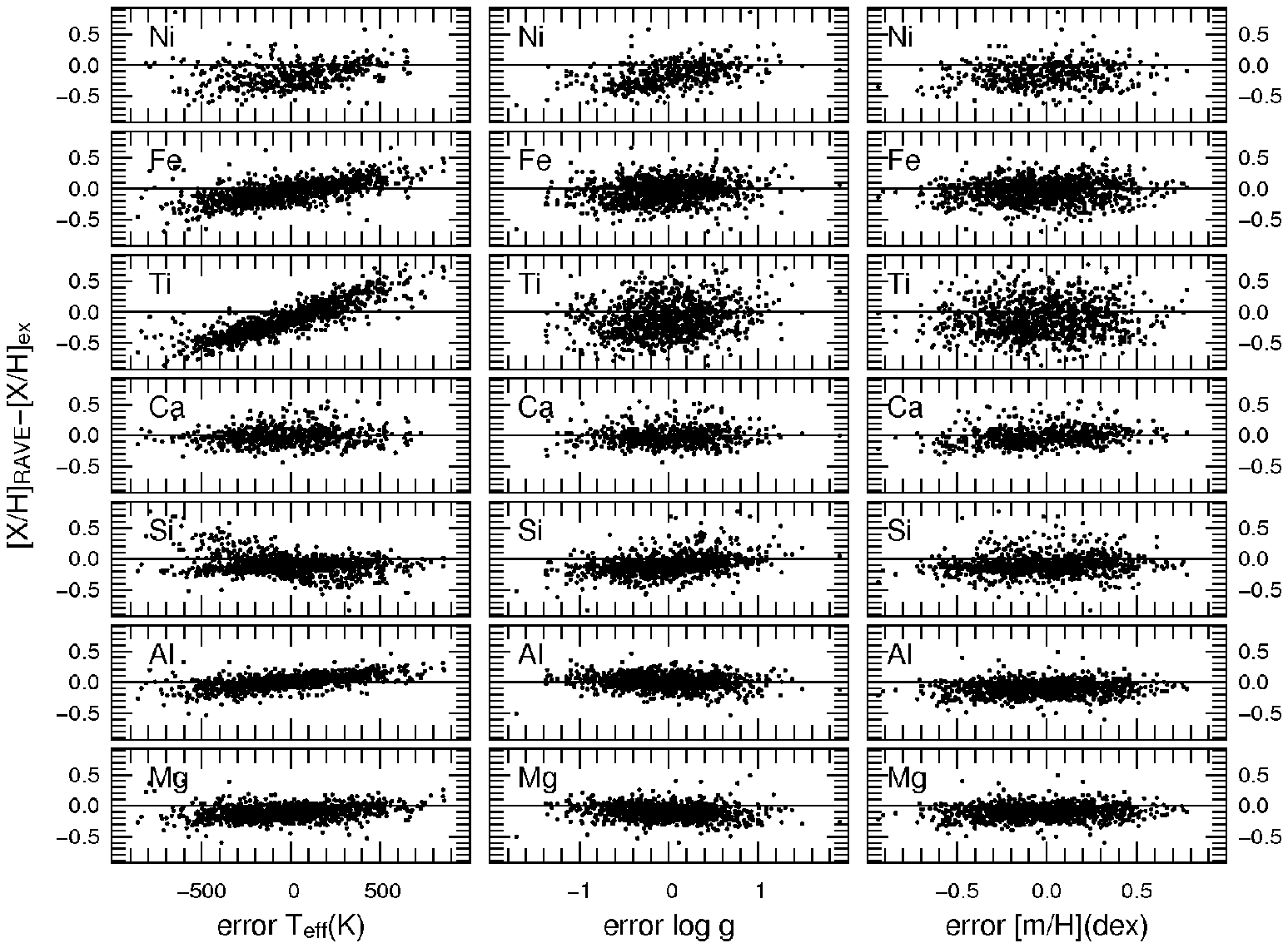}
\caption{Correlation between the elemental abundance residuals measured-minus-expected
(y-axis) and the stellar parameter errors (x-axis) at S/N=100.}
\label{abd_err_comp}
\end{figure}

As could be expected, there is a correlation between errors in stellar
parameters and errors in the elemental abundance estimates.
Figure~\ref{abd_err_comp} highlights these correlations, showing that
most of the elemental abundances errors are correlated with \temp\
errors.  In
general, an overestimate in \temp\ corresponds to an overestimate in
abundance, since most of the lines have smaller EW at higher \temp
. Only weak correlations with \logg\ and \met\ errors are evident.

\subsection{The accuracy of elemental abundance estimates from real spectra}

The initial RAVE input catalogue had no stars whose chemical
elemental abundances were known to high accuracy to compare the results of our pipeline with. 
To create a comparison dataset of real stars, we therefore observed 104 stars chosen from
Soubiran \& Girard \citep[][]{soubiran} (hereafter SG05), a collection
of high-precision elemental abundance measurements from the literature. Since
these are bright stars, most of the RAVE spectra have S/N$>$100 and therefore the
accuracy in our abundance estimates is representative of the high S/N
case. Ruchti et al. \citep[][]{ruchti} (hereafter R10) have also
re-observed 243 RAVE stars at high resolution and measured stellar
parameters and chemical elemental abundances. The RAVE spectra of these stars
have S/N between 30 and 90 and therefore we can use them to test the
accuracy of our procedure for intermediate S/N.

We have seven elements in common with SG05 and five elements with R10
(Fe, Mg, Si, Ca and Ti). However, for the latter we could only compare four elements
because the weak \Cai lines were not strong enough in the RAVE spectra
to be measured by the pipeline. As with synthetic spectra, we ran a
first test by adopting the high-resolution stellar parameters to
evaluate the performance of the pipeline alone and then ran a second
test by using the RAVE stellar parameters, which have larger
uncertainties relative to the high-resolution data.

\begin{figure}[t]
\centering
\includegraphics[width=12cm]{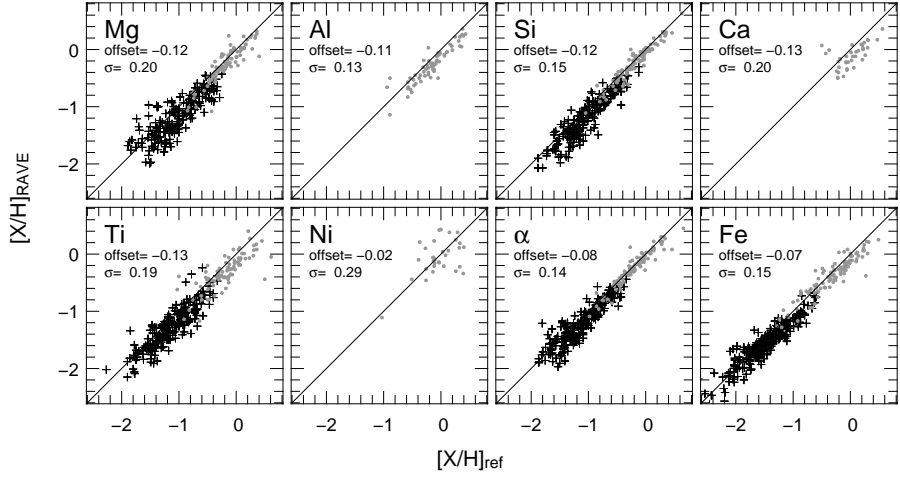}
\caption{Comparison between high resolution elemental abundance
([X/H]$_{\mbox{ref}}$) from the literature
(x-axis) and RAVE elemental abundances (y-axis) for the SG05 sample (98 stars, gray dots) and
the R10 sample (243 stars, black ``$+$''). For these measurements we adopted
the values of \temp\ and \logg\ from the high-resolution data (first test).}
\label{highres_Ruchti_X_H}
\end{figure}

\begin{figure}[b]
\centering
\includegraphics[width=12cm]{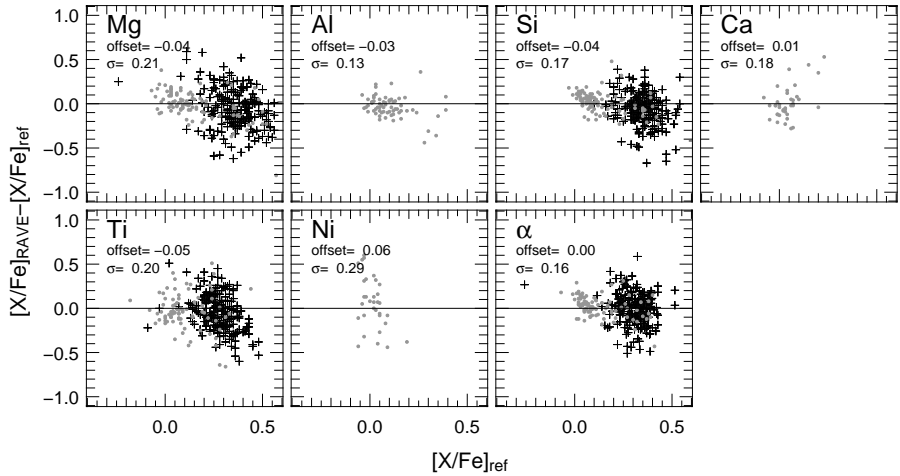}
\caption{Relative elemental abundance residuals RAVE-minus-high resolution (y-axis)
versus high resolution measurements ([X/Fe]$_{\rm ref}$). 
For these measurements we adopted
the values of \temp\ and \logg\ from the high-resolution data
(first test). The symbols are as in Figure~\ref{highres_Ruchti_X_H}.}
\label{highres_Ruchti_X_Fe}
\end{figure}

The results of the first test are shown in Figure~\ref{highres_Ruchti_X_H}
where the gray dots and black ``$+$'' correspond to SG05 and R10,
respectively.  When the high-resolution stellar parameters are adopted the
chemical pipeline's elemental abundances agree with the SG05 and R10 results
to within $\sim 0.15$ dex on average.  The smaller dispersion of the SG05
stars around the one-to-one correspondence line compared to R10 is due to
the higher S/N of the RAVE spectra of these stars.  All elements are
typically underestimated by $\sim$0.1 dex, an offset that is roughly
constant across the entire abundance range [$-2.0$,$+0.5$] dex.  The
$\alpha$-enhancement (Figure~\ref{highres_Ruchti_X_Fe}) is estimated well
for all but [$\alpha$/Fe]$>$0.4 dex, where there is a small systematic bias
of $\sim$0.1 dex.

\begin{figure}[t]
\centering
\includegraphics[width=12cm]{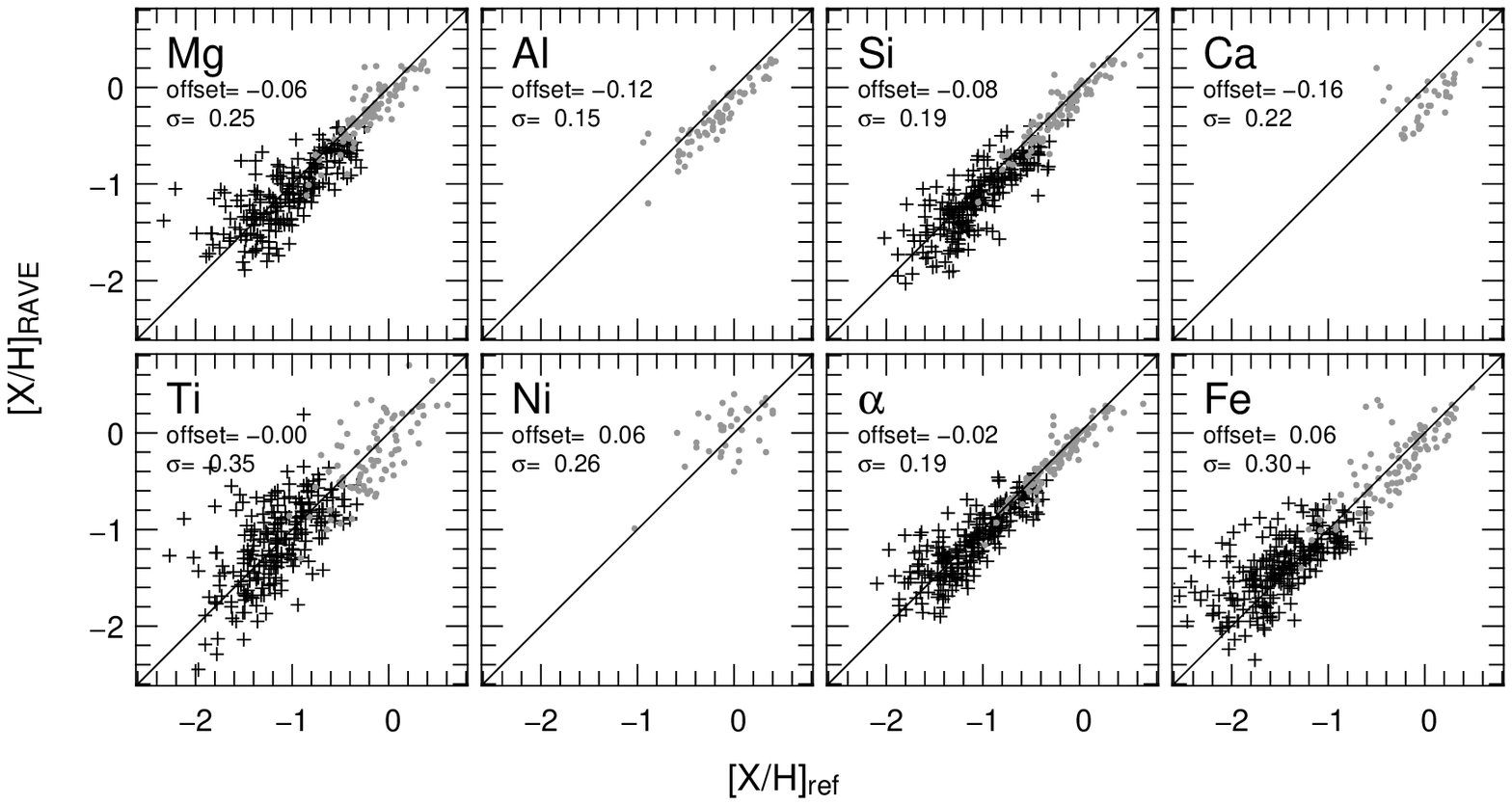}
\caption{As in Figure~\ref{highres_Ruchti_X_H} but adopting \temp\ and
\logg\ from the RAVE data (second test).}
\label{RAVE_Ruchti_X_H}
\end{figure}

When the RAVE stellar parameters are adopted (the ``second test",
Figure~\ref{RAVE_Ruchti_X_H}) the dispersion in our elemental abundances
compared to the published values increases to $\sim$0.2-0.3 dex on 
average, depending on the element. In Table~\ref{table_stat}
  we report the mean and standard deviation after grouping the sample
  in dwarf and giant stars separately.  The larger dispersion at low
abundances ([X/H]$<$--1.0 dex) is very likely due to the fact that
estimating precise stellar parameters on such spectra is challenging
due to the few and weak visible absorption features (excluding for the
\Caii triplet, which is not measured). 

\begin{figure}[b]
\centering
\includegraphics[width=12cm]{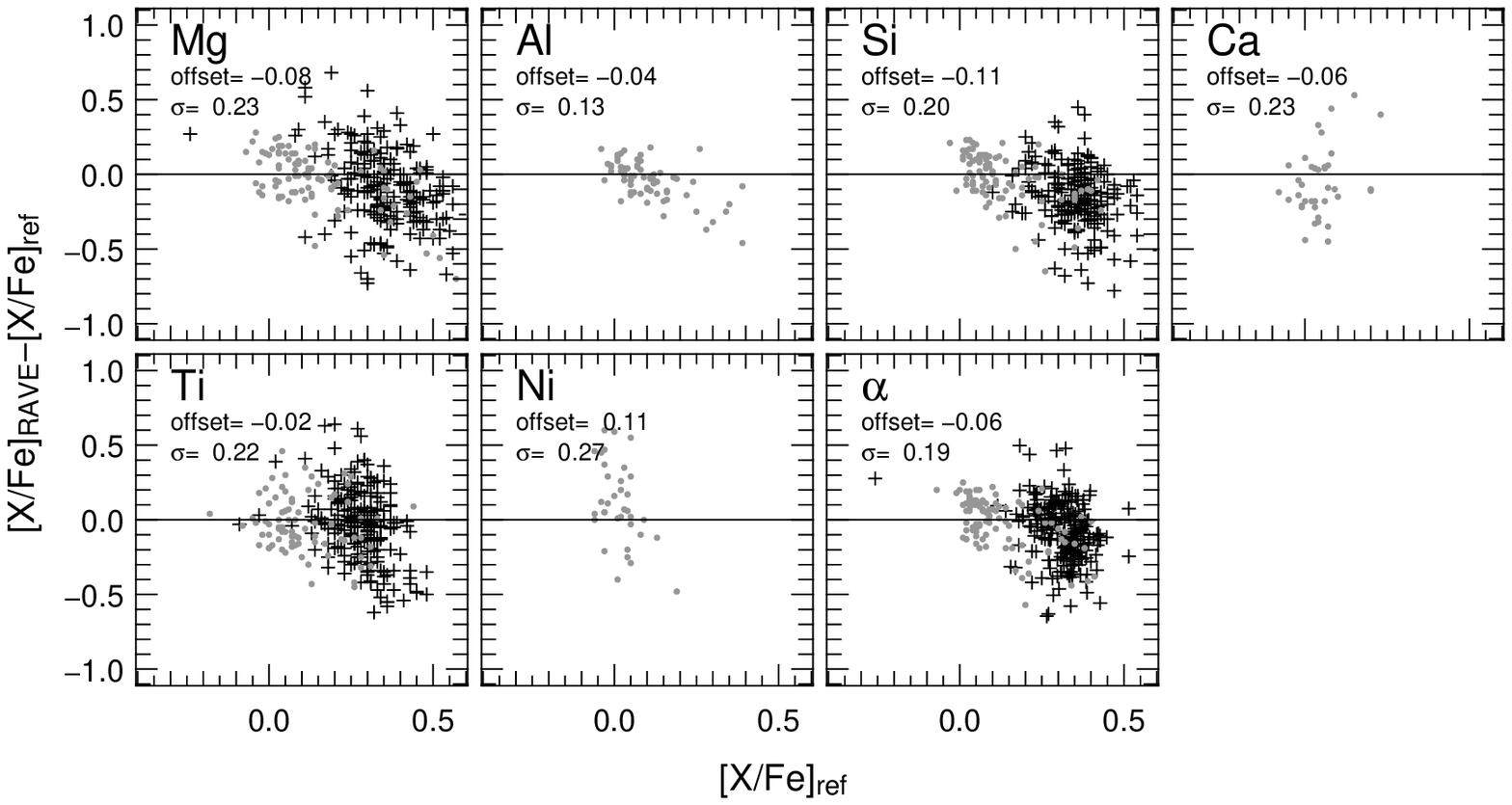}
\caption{As in Figure~\ref{highres_Ruchti_X_Fe} but adopting \temp\ and
\logg\ from the RAVE data (second test).}
\label{RAVE_Ruchti_X_Fe}
\end{figure}

The enhancement estimates ([X/Fe], see
Figure~\ref{RAVE_Ruchti_X_Fe}) are accurate to within 
$\sim$0.3 dex. At high enhancements, abundances are systematically
underestimated by $\sim$0.1-0.2 dex, depending on the element.  However, as
we are comparing our measurements with other authors' (high resolution)
measurements, the variance of the residuals
is the quadratic sum of the variances of our and the high resolution
results. This means that the standard deviations reported in
Figure~\ref{RAVE_Ruchti_X_H} and Figure~\ref{RAVE_Ruchti_X_Fe}
represent a conservative estimate of the expected RAVE errors.

\begin{figure*}[t]
\begin{minipage}[t]{9cm}
\includegraphics[width=9cm]{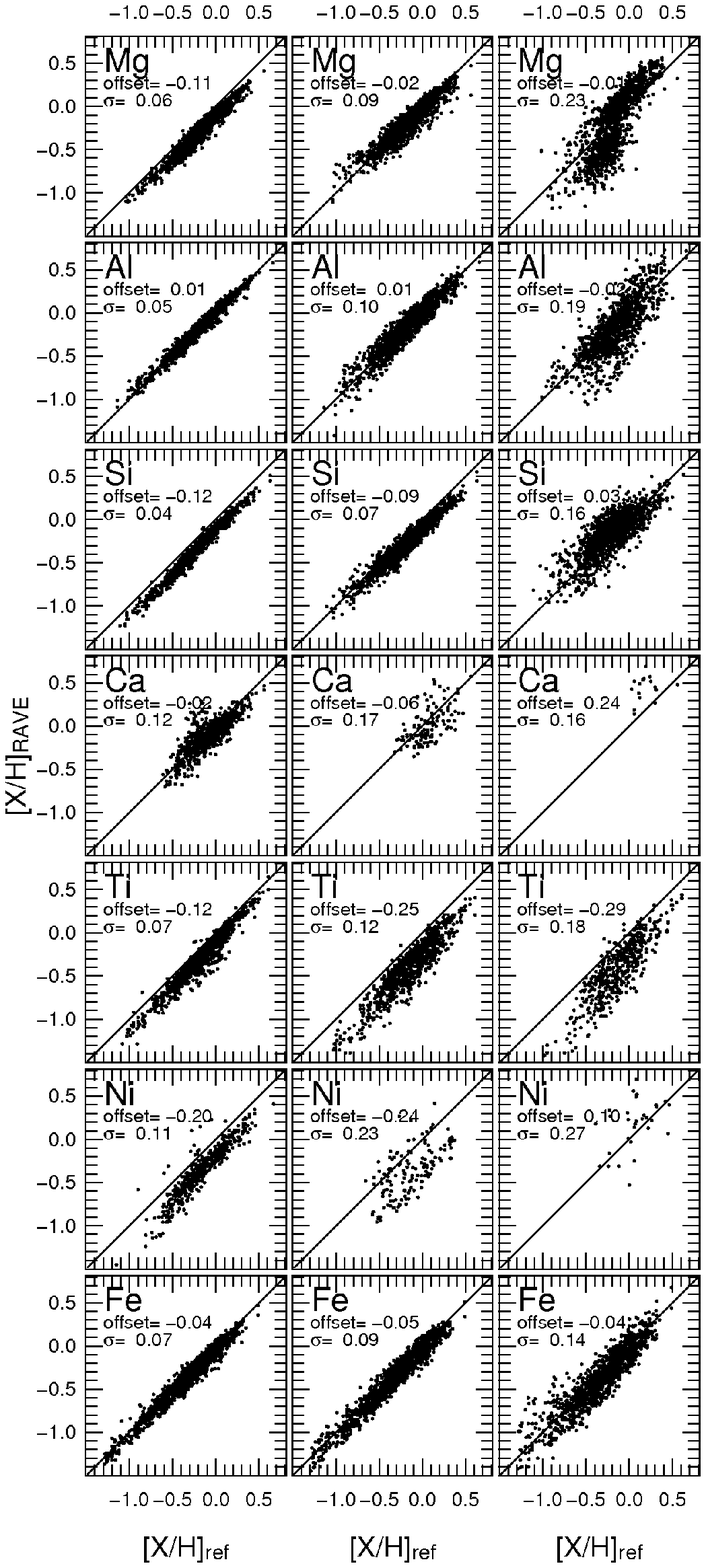}
\end{minipage}
\hfill
\begin{minipage}[t]{9cm}
\includegraphics[width=9cm]{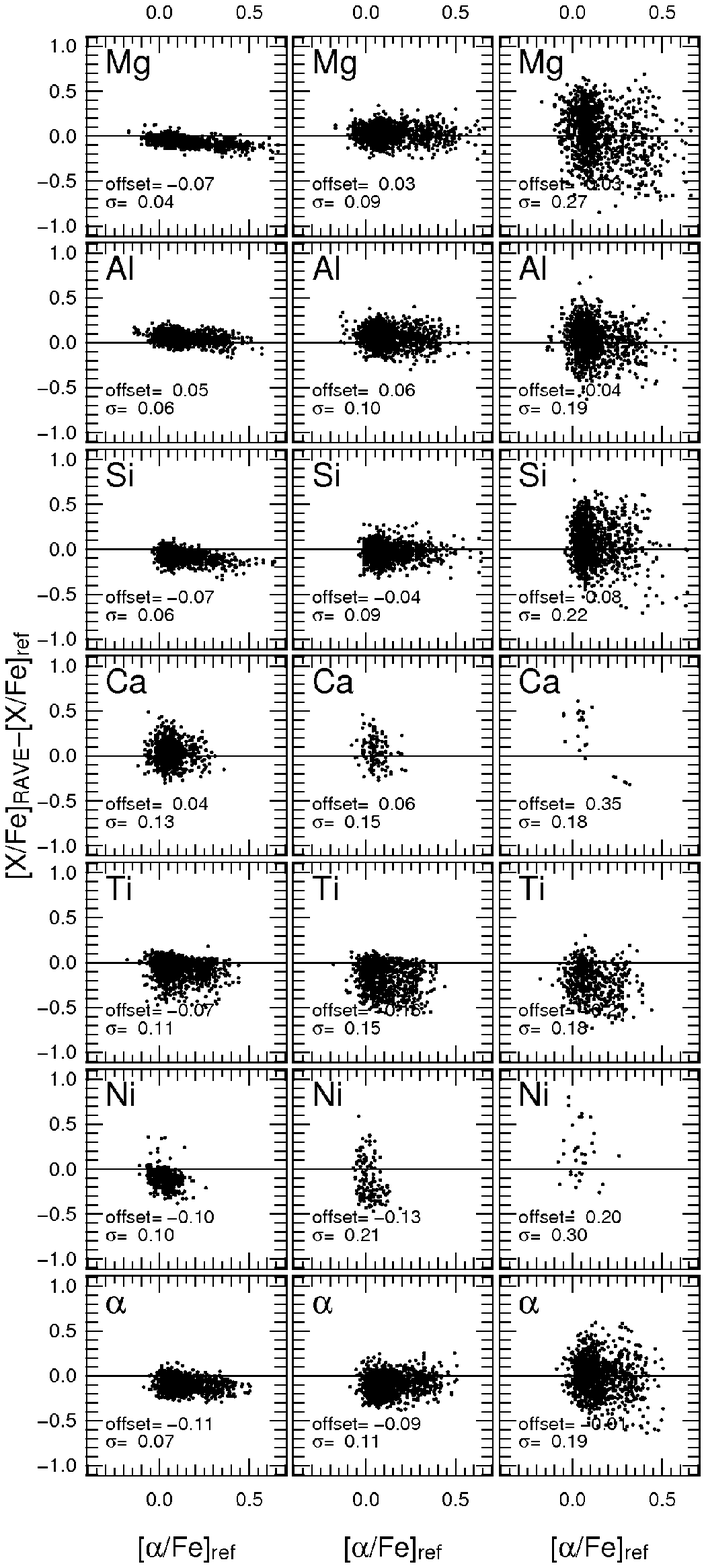}
\end{minipage}
\caption{{\bf Left}: expected elemental abundances [X/H] (x-axis) versus measured elemental abundances
(y-axis) for the sample of
synthetic spectra at S/N=100, 40, 20 (for the left, middle and right column,
respectively) and assuming no errors in stellar parameters.
{\bf Right}: as in left panels
but for the expected enhancement [X/Fe] (x-axis) and the residuals
measured-minus-expected (y-axis). Offsets and standard deviations are
reported in the panels.}
\label{abd_comparison}
\end{figure*}

\begin{figure*}[t]
\begin{minipage}[t]{9cm}
\includegraphics[width=9cm]{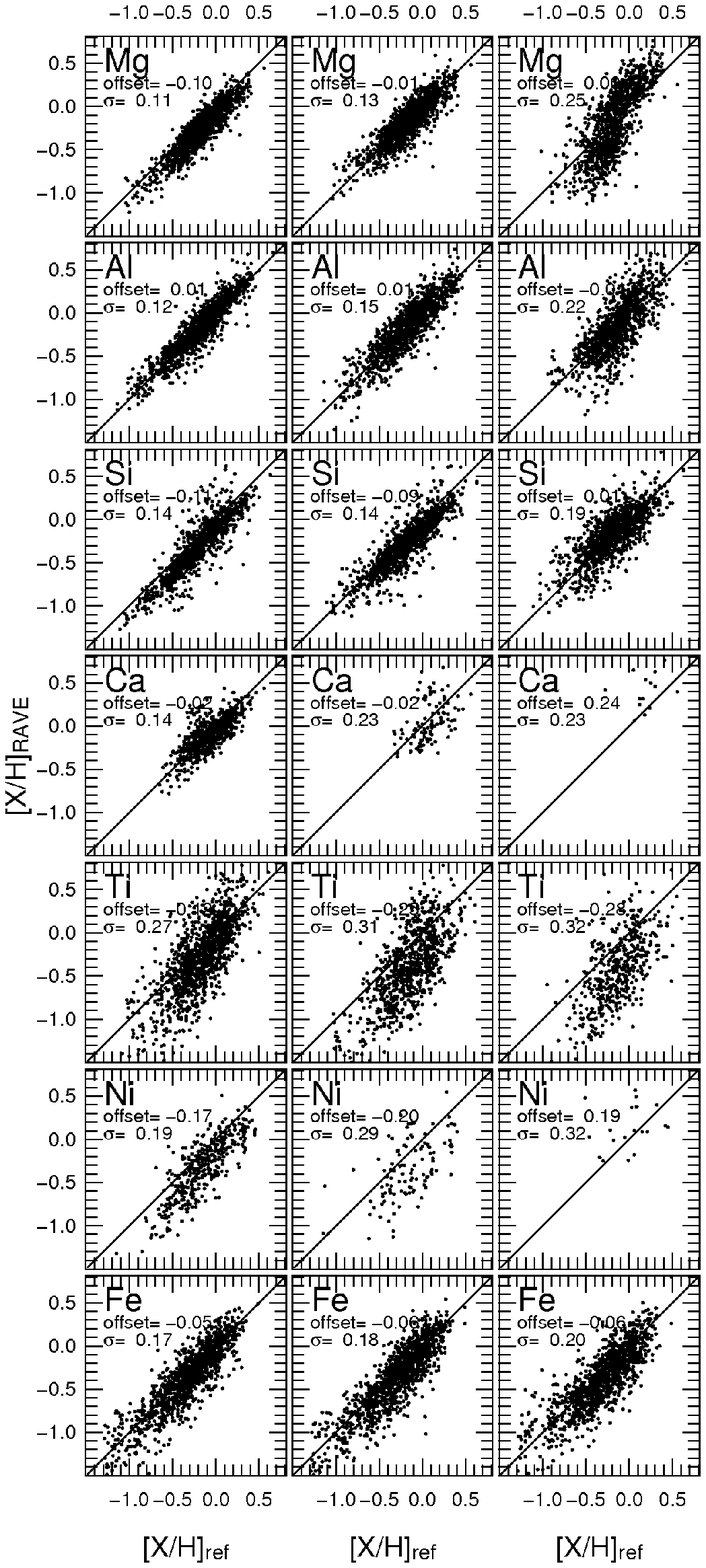}
\end{minipage}
\hfill
\begin{minipage}[t]{9cm}
\includegraphics[width=9cm]{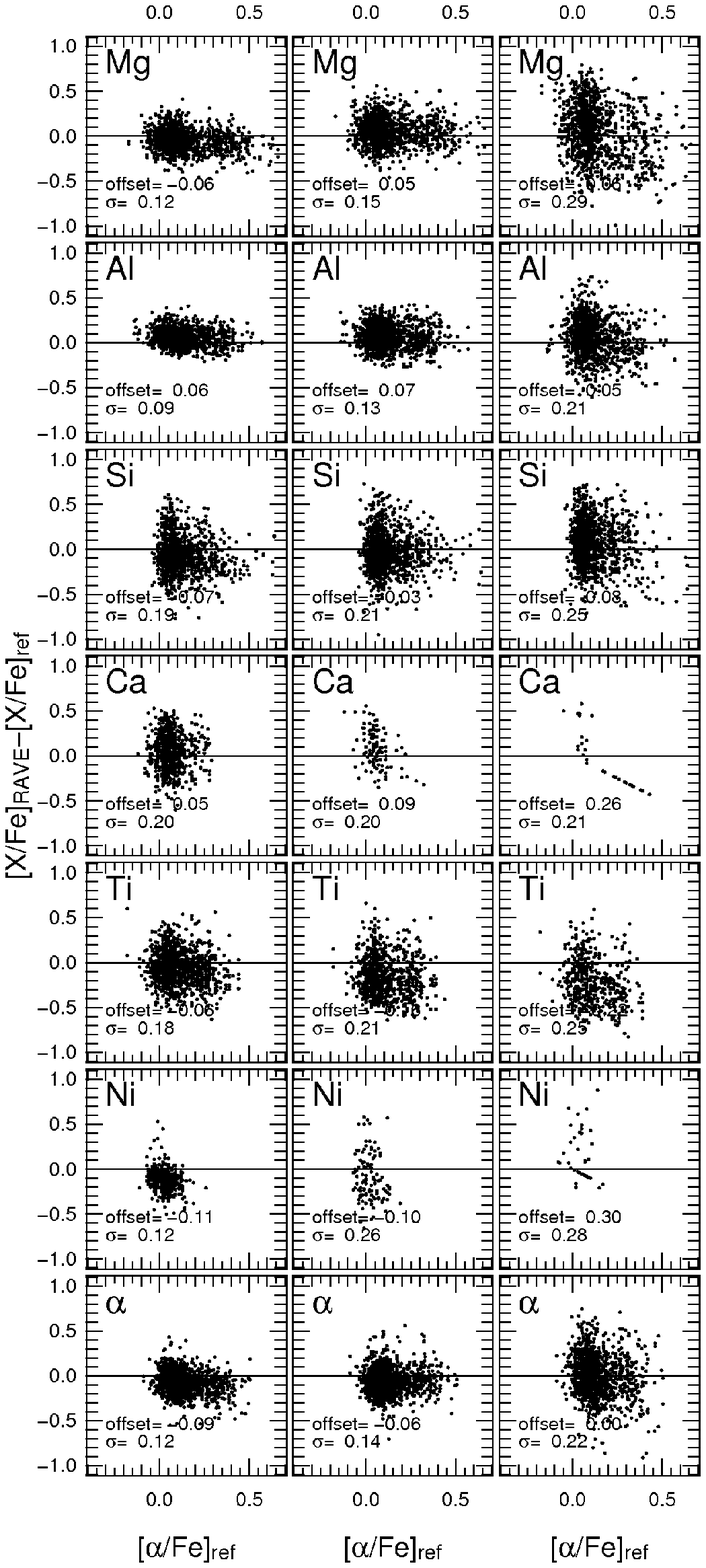}
\end{minipage}
\caption{As in Figure~\ref{abd_comparison} but with noisy stellar parameters to simulate
the RAVE stellar parameters of the RAVE archive.}
\label{abd_comparison_err}
\end{figure*}

\begin{table}[t]
\begin{center}

\begin{tabular}{l|rr|rr||rr|rr}
\tableline\tableline
&\multicolumn{4}{c}{high res. stellar parameters}&\multicolumn{4}{c}{RAVE stellar parameters}\\
 &\multicolumn{2}{c}{giants}& \multicolumn{2}{c}{dwarfs}&\multicolumn{2}{c}{giants}&
\multicolumn{2}{c}{dwarfs}\\
 & mean & $\sigma$(dex) & mean& $\sigma$(dex)& mean & $\sigma$(dex) & mean& $\sigma$(dex)\\ 
\tableline
$[$Mg/H$]-[$Mg/H$]_{\mbox{ex}}$ &-0.15  &0.21  & -0.08  &0.14   &-0.05  &0.28  &-0.06  & 0.15 \\
$[$Al/H$]-[$Al/H$]_{\mbox{ex}}$ &  & &-0.11   &0.13   &  &  &-0.11  &0.15  \\
$[$Si/H$]-[$Si/H$]_{\mbox{ex}}$ &-0.15  &0.15  &-0.04   &0.09   &0.11  &0.21  &-0.03  &0.14  \\
$[$Ca/H$]-[$Ca/H$]_{\mbox{ex}}$ &  &  &-0.13   &0.20   &  &  &-0.16  &0.22  \\
$[$Ti/H$]-[$Ti/H$]_{\mbox{ex}}$&-0.14 & 0.17 & -0.12 & 0.22 & 0.10 &  0.36& -0.03 & 0.30 \\
$[$Fe/H$]-[$Fe/H$]_{\mbox{ex}}$ & -0.08 &0.15  &-0.06   &0.15   & 0.07 & 0.30 & 0.04 & 0.31 \\
$[$Ni/H$]-[$Ni/H$]_{\mbox{ex}}$ &  &  &-0.02   & 0.29   &  &  &0.06  &0.26  \\
$[\alpha$/H$]-[\alpha$/H$]_{\mbox{ex}}$ &-0.10  &0.15  & -0.04  &0.10   & -0.02  & 0.21  & -0.02 & 0.15 \\
\tableline
\end{tabular}
\caption{Mean and standard deviation of the residuals between measured and
expected elemental abundances obtained by using the high resolution stellar parameters
(left) and the RAVE parameters (right) for 347 standard stars. 
We computed the values for
giants (\logg$<$3.5) and dwarfs (\logg$\ge$3.5) separately. The two groups
are also representative of stars with \temp$<$5500K and intermediate STN
(R10 stars), and stars with \temp$>$5500K and high STN (SG05 stars).
}
\label{table_stat}
\end{center}
\end{table}

\begin{figure}[b]
\centering
\includegraphics[width=12cm]{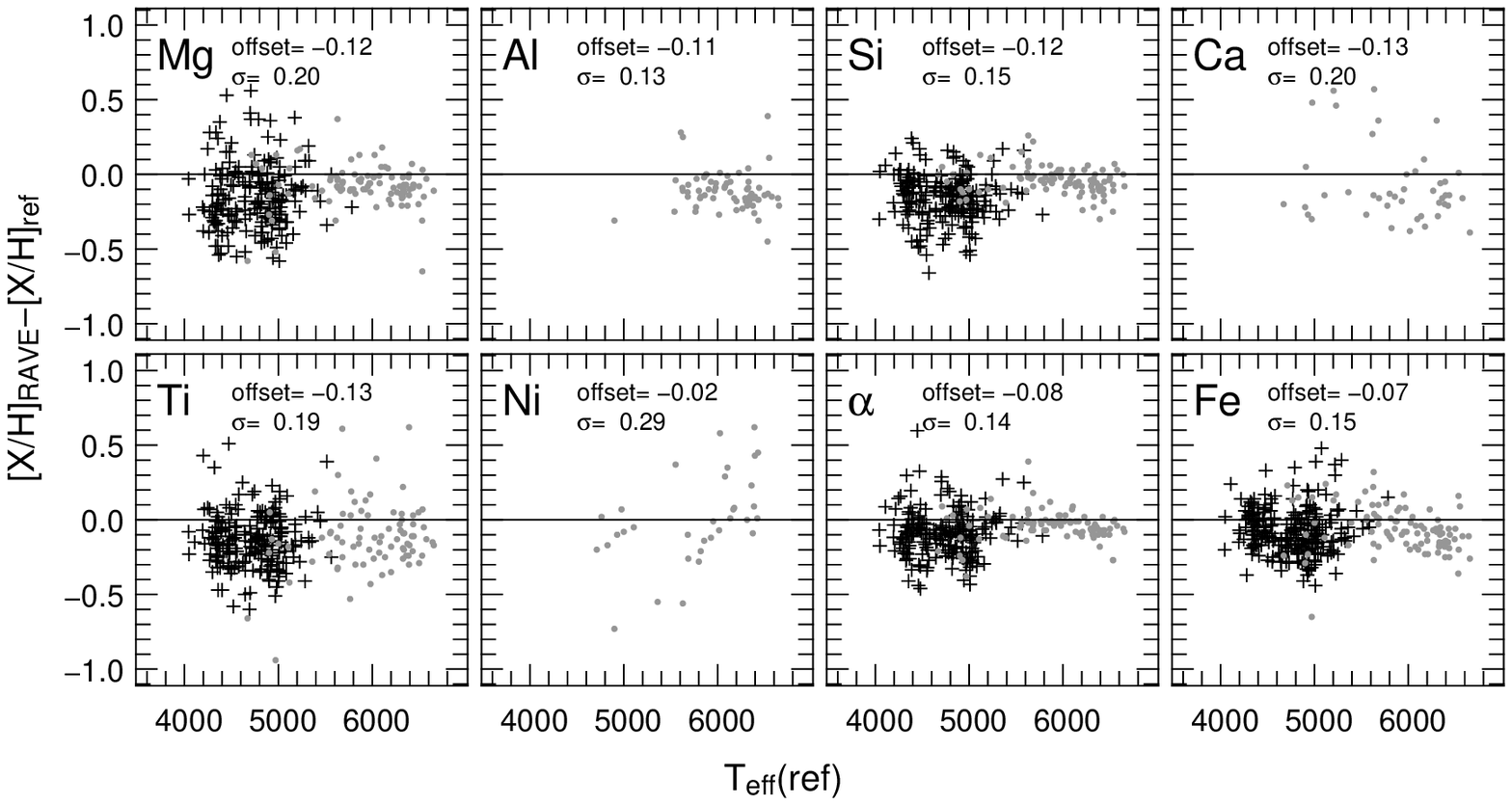}
\caption{Residuals between measured and expected elemental abundances as a function of \temp .
For these measurements we adopted high resolution stellar parameters.}
\label{resid_abd_Teff}
\end{figure}

\subsection{Internal errors from repeat observations}

\begin{figure}[t]
\centering
\includegraphics[width=12cm]{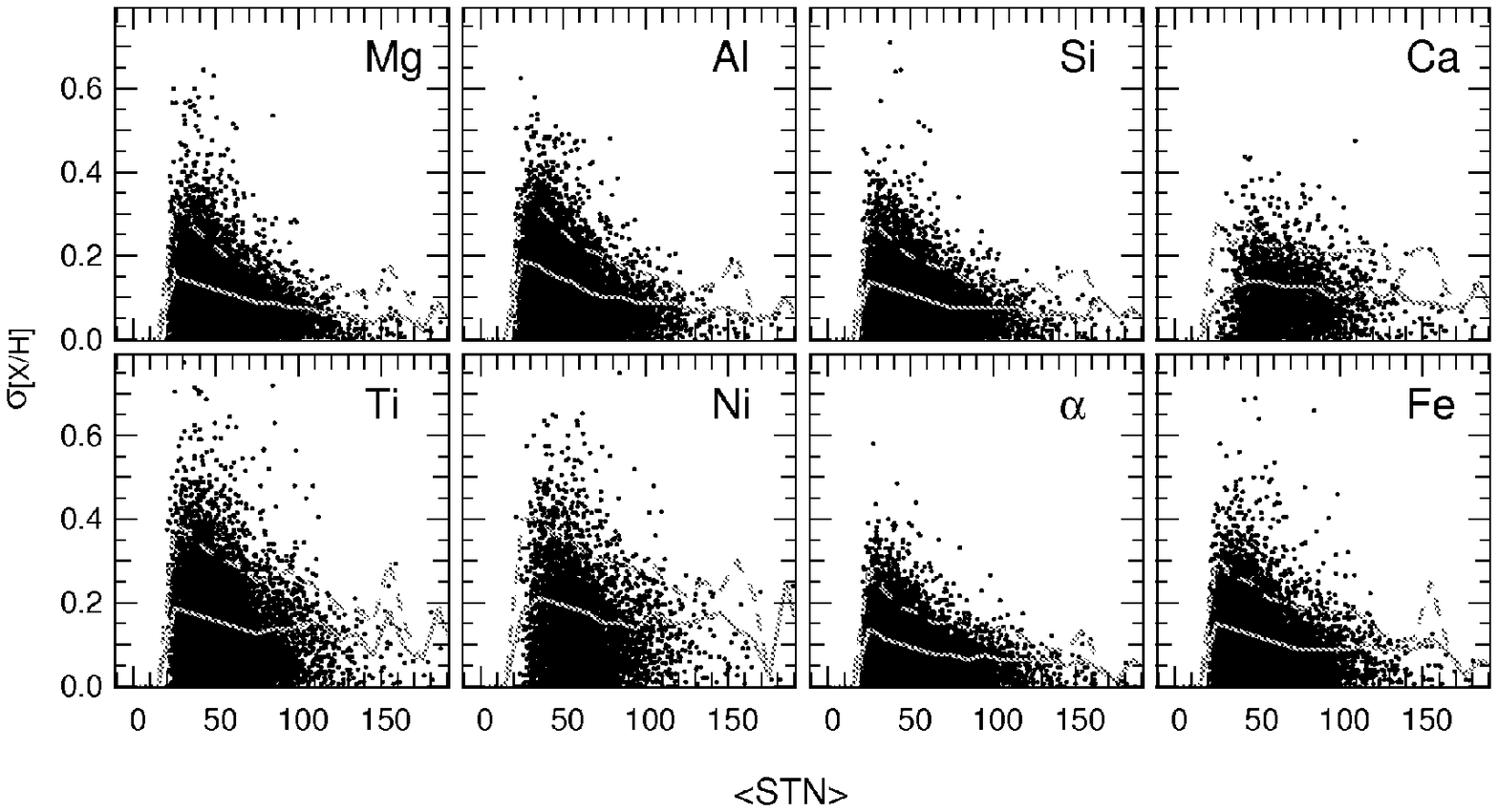}
\caption{Expected internal errors of the chemical elemental abundances, estimated via
standard deviations of the abundances from repeat observations. 
For a more robust statistic we used 30821 spectra of 12504
stars of the RAVE internal data release. The solid gray line represents the
1$\sigma$ abundance error limit, i.e., 68\% of the stars lie under the line
for each STN bin, 
whereas the dashed gray line represents the 2$\sigma$ limit. For this
sample 9469 stars have two observations, 1286 have three, 1060 have four, 473
have five, 153 have six, 33 have seven, 11 have eight, 11 have nine, 4 have
ten and four have 11 observations.}
\label{repetitions}
\end{figure}

In order to take into account the effect of binary stars, RAVE
observes some stars ($\sim$5\% of the whole sample) more than once.
In this chemical catalogue, $964$ stars have been observed twice,
$90$ thrice and $67$ four times. Multiple observations are also
useful to estimate the internal error by comparing the results
obtained from different spectra of the same object.  A fair
comparison would require that the repeated observations of an object
have the same S/N. Unfortunately, due to variable atmospheric
conditions the spectra often have different S/N values.  We
therefore averaged the STNs of the spectra for each object and
computed the corresponding standard deviations of the elemental abundances. In this way we could assess
the magnitude of internal errors in our procedure as a function of STN. 
In Figure~\ref{repetitions} we plot the results for $12,504$ RAVE stars having multiple
observations, where we use the RAVE internal data release as it has a substantially larger number
of stars. For all the elements the 1$\sigma$ confidence line (gray solid line) lies below
0.2 dex and hovers around $\sim$0.1 dex for STN$>$80.

\subsection{\Mrave\ vs. \Mchem: a comparison}

\begin{figure}[t]
\centering
\includegraphics[width=14cm]{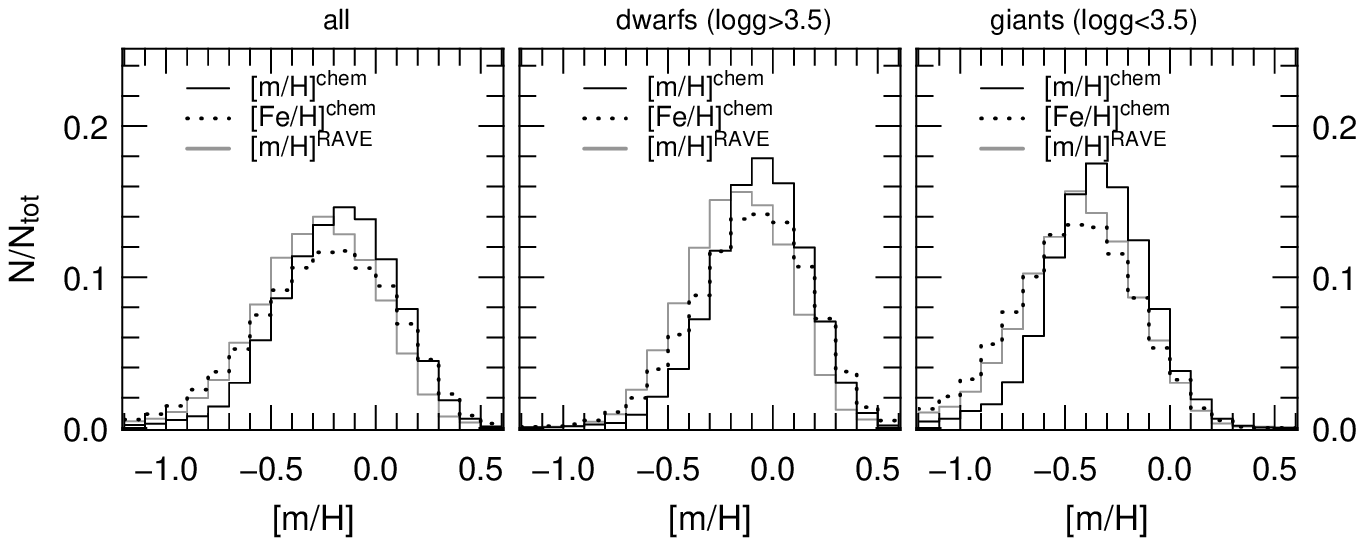}
\caption{Comparison between \Mchem\ (black line) \Fechem\ (dotted line) and
\Mrave\ (non-calibrated metallicity, gray line) 
distributions for the whole sample (37848 spectra, left panel), dwarf (middle panel) and
giant stars (right panel). }
\label{M_rave_chem}
\end{figure}

In Figure~\ref{M_rave_chem} we compare the distributions of \Mrave, 
\Mchem\  and \Fechem\ for 37819 RAVE spectra with STN$>$20.
The metallicity \Mchem\ is inferred from the chemical elemental abundances with the
equation given by Salaris et al. \citep[][]{salaris}
\begin{equation}\label{eq_rave_met}
[m/H]^{chem}=\mbox{[Fe/H]}+ \log (0.638\cdot10^{\mbox{[$\alpha$/Fe]}}+0.362),
\end{equation}
where the $\alpha$ enhancement is computed as
\begin{equation}\label{alpha_enh}
\mbox{[$\alpha$/Fe]}=\frac{\mbox{[Mg/H]+[Si/H]}}{2}-\mbox{[Fe/H]}
\end{equation} 
and the elemental abundances [Mg/H], [Si/H] and [Fe/H] come from the chemical
pipeline.

\Mrave\ is $\sim$0.1 dex lower than \Mchem, which is to be expected as
the non-calibrated RAVE metallicity is
underestimated by $\sim$0.15 dex with respect to the reference stars
used in \citealp[Siebert et al.,][]{siebert}. The shape of the  \Mrave\ distribution
is fairly similar to the \Mchem\ distribution for dwarf stars
but different for giants. Moreover, \Mrave\ seems to better match
\Fechem\ than \Mchem, particularly for giants. 
This could be due to $\alpha$ enhancement: giants stars have a higher 
proportion of thick disk stars ($\alpha$-enhanced) than 
do dwarfs stars, which are mostly thin disk (not
$\alpha$-enhanced). 
Although used during the stellar parameters estimation process, the RAVE pipeline 
is unable to measure $\alpha$ enhancement \citep[Zwitter et al.,][]{zwitter}. Therefore, it is 
possible that particularly for $\alpha$-enhanced
stars \Mrave\ better represents [Fe/H]. 
Apart from the discussed shift of 0.1 dex in
metallicity, there seems to be fair agreement between
the \Mrave\ and \Mchem\ distributions.

\section{Discussion}\label{discuss_sec}

For each element, the accuracy of the abundance estimate depends on the number of lines that are strong enough to be clearly
detected; to be measurable, the lines must be
stronger than the noise. At the same time, the intensity of the lines depends on the
stellar parameters and elemental abundances of the star. This means that for a
fixed S/N, the accuracy of the elemental abundance is a function of the
stellar parameters and of the abundance itself.

The interplay of these factors makes the accuracy of the abundance results difficult to
estimate.  At high S/N, the accuracy varies from element to element, 
depending on the number and intensity of the lines measured. For
instance, Ti and Ni lines are measured better in stars with \temp\
lower than $\sim$5000K, whereas sulfur (not present in this
data release) can be seen and measured only for temperatures higher than
$\sim$4800K. The RAVE line list has a few weak \Cai lines that can be
measured only in stars with \met\ $>$--0.5 dex.

Noise affects the EWs of the lines and
so the model spectrum. As the pipeline varies the abundances
of the individual elements, the larger the number the absoption lines of an
element, the lower the abundance error due to the fitting, because the effects of noise are averaged
over the lines. Thus, the abundance accuracy of elements with fewer lines 
is especially affected by noise. When the noise is particularly strong with respect to
the intensity of the lines, i.e.,  low S/N, the minimization routine
fits the noise and the resulting elemental abundance diverges. In such cases
the pipeline rejects the measurement and renders a null value -9.99.

The situation get worse in the low S/N regime, because the
pipeline measures even fewer lines, i.e., only those strong enough
to be detected despite the noise. This decreases the number of
measurable elements and increases the uncertainties. In
Section~\ref{test_synt} we tested the pipeline down to S/N=20 to see
if at such a low S/N the measurements are still trustworthy. The
results suggest we can use such data, but with care. Noise
generates selection effects: spectra having low \met\ or high \temp\
do not have lines strong enough to overcome the noise and they go
through the pipeline unmeasured. A limited number of lines can also
generate systematic errors because the abundance of an element depends
on just a few lines (sometimes even only one) and thus becomes
especially vulnerable to uncertainties in the oscillator strengths.

Nevertheless, the tests performed with synthetic spectra at S/N=20
show that abundances of elements with strong lines like Fe, Al and
Si can still be measured. At such low S/N, errors as large as 
$\sim$0.2-0.3 dex are expected, but they do not show large systematic biases
and the residuals are on average close to zero.

We conclude that elemental abundances may be trusted down to S/N=40 for seven
elements (details will follow in Section~\ref{sec_catalogue}), whereas
between S/N=20 and 40 we can trust the abundance of [Fe/H] and (to be
on the safe side) the abundance [$\alpha$/H], when computed from the
average of Mg and Si. Abundances for other elements for the low S/N stars
should be considered indicative, and used only for exploratory
purposes until new comparison stars enable a proper validation of our
results in this S/N regime.

\subsection{Zero point of the RAVE elemental abundance scale} 

Our elemental abundance measurements are indirect measurements in the sense that
they are inferred from the comparison between the intensity of lines
seen in real spectra and their intensity predicted by stellar
atmosphere models. Since the models are different for different
stellar parameters, a question arises regarding whether all the models
yield elemental abundances which refer to the same zero point, i.e., the origin
of the internal elemental abundance scale. The same question concerns whether
this zero point refers to the same zero point of the real spectra, i.e.,
between the internal and external scales.

The latter question has a prompt answer: we do not know the external
scale, because real stellar atmospheres have never been directly probed and all
elemental abundance measurements refer to models. Therefore, we can only
check the consistency of the internal scale. This can be performed by
comparing the measured elemental abundances of a sample of synthetic spectra at
different stellar parameters, as done in Figure~\ref{abd_comparison}.

In this plot, at S/N=100, the residuals between measured and expected
elemental abundances are on average zero for any gravity and metallicity; the points align along a
straight line with slope roughly equal to one. This means that the
measured differences in elemental abundance at any metallicity regime are the
same (constant offset), i.e., they refer to the same zero point. However, the
offset from element to element may vary, showing that the offsets are due 
to the measurement process and not
due to the metallicity of the atmosphere models.

On the other hand, there is a clear correlation between residuals and
\temp : the higher the \temp\ ,the higher the measured elemental abundance.  This
systematic may be due to continuum correction: cooler stars have spectra that are crowded with
absorption lines and so the continuum appears lower than it is. Additionally, the
varying behaviour of the elements in
Figure~\ref{abd_comparison} can be explained by considering that their
lines lie in different regions of the spectrum, i.e., on the wings of a
\Caii line or in a region relatively free of lines. This can translate into 
systematic, wavelength-localized shifts of the estimated
continuum as a function of \temp .

Although this systematic effect appear tractable and correctable, we did not apply any correction to the data because such a
systematic is not visible on tests with real spectra. When the RAVE
elemental abundances are compared with the expected elemental abundances of standard stars
as function of \temp\, no clear trend is visible
(Figure~\ref{resid_abd_Teff}), except that RAVE elemental abundances 
have a slight systematic bias to lower values. 
A reasonable explanation is that the elemental abundance measurements 
given by SG05 and R10 suffer from the same
systematic error, due to uncertainties in estimating the correct continuum level
in regions crowded with absorption lines or affected by the wings of strong,
broad lines.

We conclude that the RAVE chemical pipeline can determine 
chemical elemental abundances with a systematic error that may be as large as $\sim0.1$ 
dex as a function of \temp. 
Since this systematic looks quite linear with \temp\ (see Figure~\ref{abd_err_comp}
and Figure~\ref{resid_abd_Teff}), to reduce this error to $\pm$0.05 dex we suggest
analysing separately samples of stars with \temp$>$5500K (mostly
dwarfs) from samples with \temp\ $<$5500K (mostly giants). 

\section{The RAVE chemical catalogue}\label{sec_catalogue}
We present the catalogue of chemical elemental abundances for 36,561 RAVE stars,
measured using 37,848 spectra.

\subsection{Sample selection}
The spectra have been selected from the DR3 RAVE database using the
following constraints:
\begin{itemize}
\item {\bf Effective temperature 4000$\leq$\temp (K)$\leq$7000}: this
  is the temperature range within which the RAVE line list has been
  calibrated.  At lower temperature the spectra are characterized by
  molecular lines other than CN (CH, TiO and other molecules), while 
  at higher temperatures lines of ionized atoms appear, and both are not
  included in the line list.
\item {\bf Signal-to-noise STN$>$20}: for STN$<$20 the absorption
  lines are strongly affected by noise and the stellar parameters and
  chemical elemental abundances are not reliable.
\item {\bf Rotational velocity V$_{rot}<$50 km s$^{-1}$}: at higher
  rotational velocity the lines have a FWHM larger than that due to
  the spectral resolution (FWHM$\simeq$1.2\AA) and they cannot be
  precisely measured. Moreover, any spectra showing larger lines
  might be a double-lined spectrum, which we also must be avoided.

\end{itemize}

Despite these selection criteria, some spectra exhibit emission or
are affected by bad continuum normalization. For such spectra the
stellar parameters and chemical elemental abundances are not reliable and it is
advisable to reject them.  Constraints on the parameters $\chi^2$ and
{\it frac} help identify these spectra. For statistical studies we
suggest rejecting spectra with $\chi^2>2000$ and {\it frac} $<$ 0.7, 
which removes 739 out of 37,848 spectra of the catalogue.

\begin{figure}[t]
\centering 
\includegraphics[width=10cm,clip]{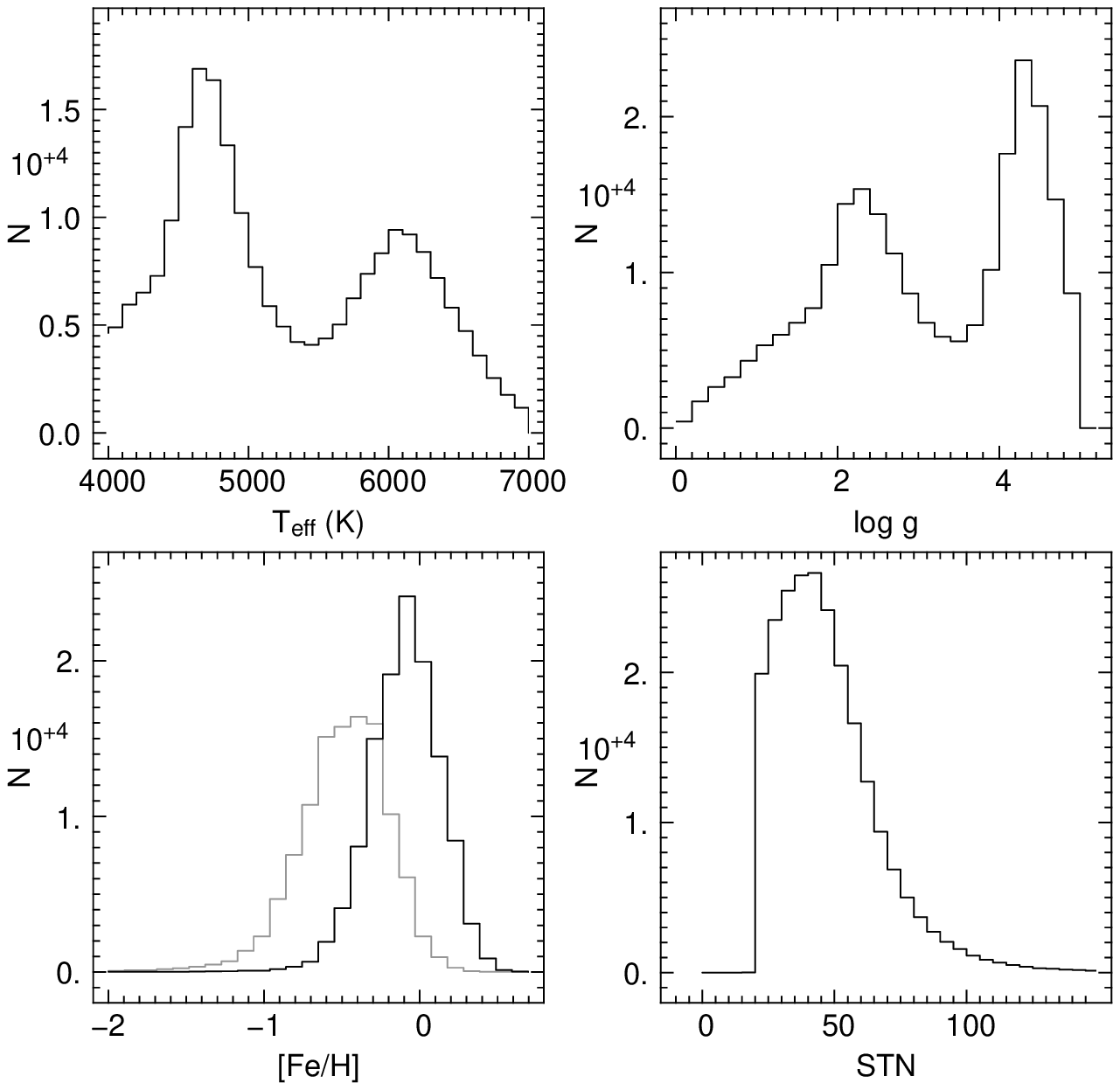}
\caption{Distributions of stellar parameters \temp ,
\logg , [Fe/H] and signal-to-noise ratio STN for 
37848 spectra of the chemical catalogue.
In the bottom left panel, the [Fe/H] distribution is given separately for dwarfs (black line) 
and giants (gray line).
}
\label{distr_s2n_T_logg_M} 
\end{figure}

\subsection{Stellar parameters}

The distributions of the stellar parameters and STN values are presented in
Figure~\ref{distr_s2n_T_logg_M}. The quantities \temp\ and \logg\ are
given by the RAVE data archive whereas STN and [Fe/H] are estimated by
the chemical pipeline. The distributions in \temp\ and \logg\ show
two peaks corresponding to giants at lower temperature and
dwarfs, which are mostly at higher temperatures. The iron abundance
distribution peaks at [Fe/H]$\simeq$--0.1 dex for dwarfs
and [Fe/H]$\simeq$--0.5 dex for giants. The latter are in average more metal-poor
because they lie further from the Galactic plane and have therefore a larger
fraction of thick disk stars\footnote{As RAVE is a magnitude limited
survey, the intrinsically bright objects are on average at large distances.}.

\subsection{Chemical elemental abundances: selection effects due to STN and stellar
parameters}

\begin{figure}[b]
\centering 
\includegraphics[bb=100 291 590 554,width=12cm,clip]{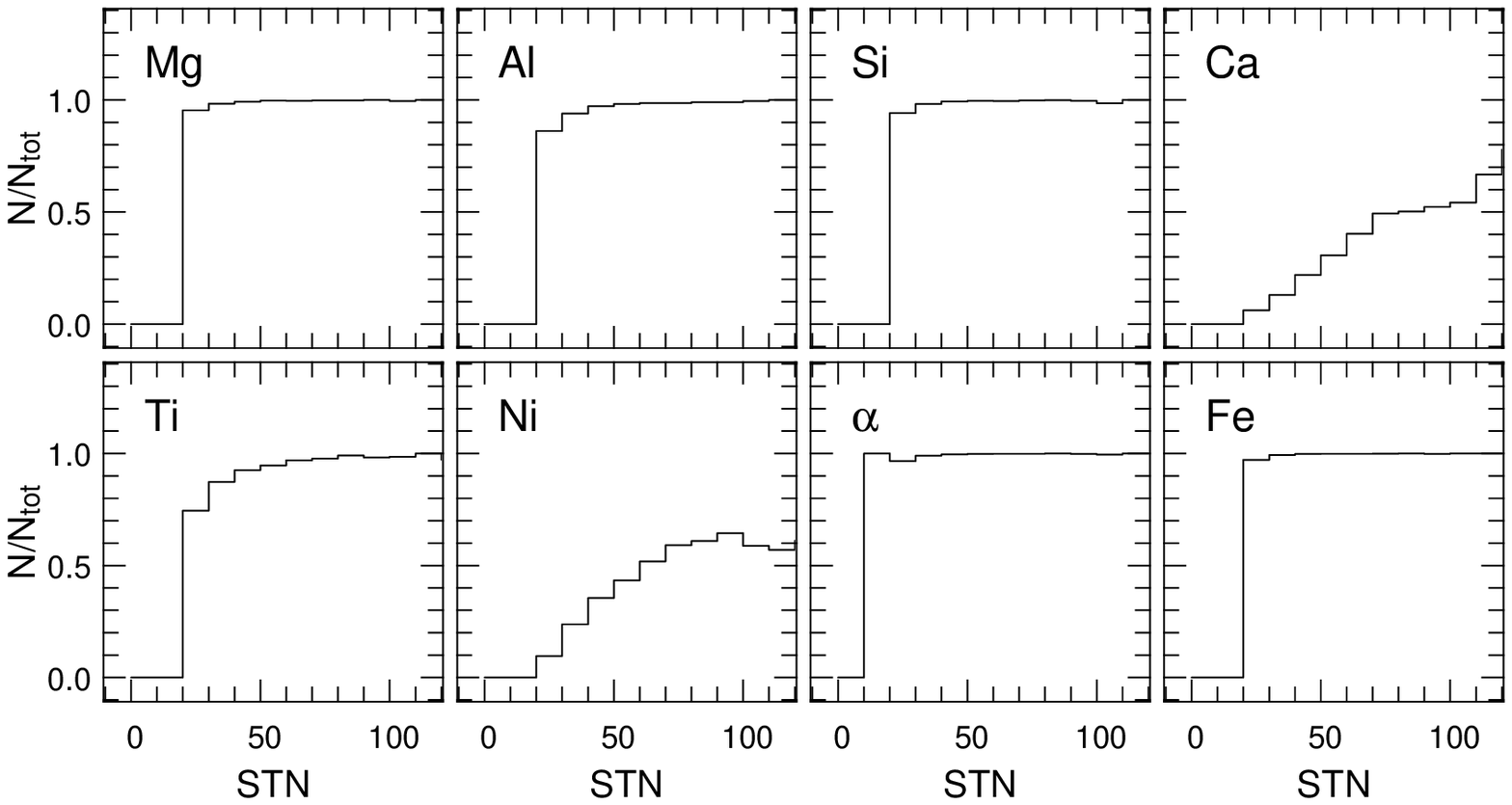}
\caption{Fraction of spectra having elemental abundance measurements as a function of the STN. 
Every bin is normalized to the total number of observed spectra in the
corresponding STN bin.}
\label{distr_stn_Nspec} 
\end{figure}

\begin{figure}[t]
\centering 
\includegraphics[bb=100 291 590 554,width=12cm,clip]{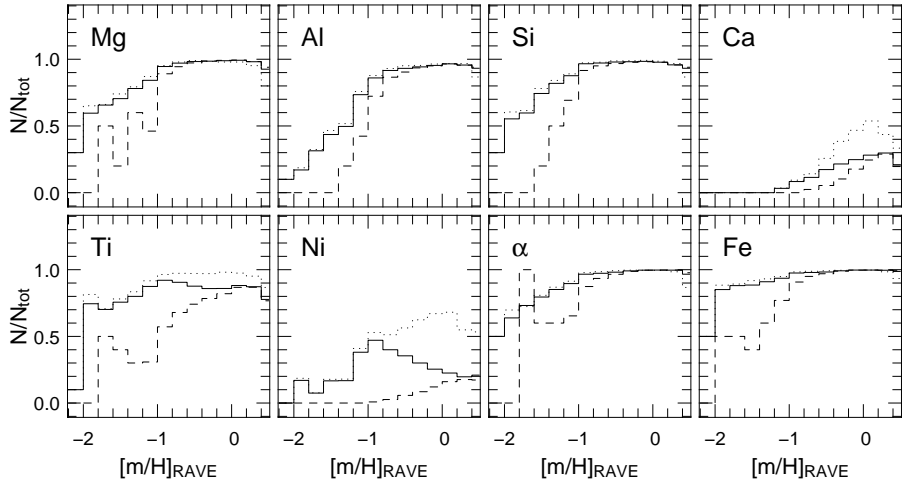}
\caption{Fraction of spectra having elemental abundance measurements as a function of
\Mrave . Every bin is normalized to the total number of observed spectra in that
\Mrave\ bin. The solid, dashed and dotted lines represent the whole sample,
dwarf stars only and giants stars only, respectively.}
\label{distr_met_Nspec} 
\end{figure}

We encourage the user to pay particular attention to selection effects
introduced by the stellar parameters and STN.
Together they affect the total number of abundance estimations as well as their accuracy.
Absorption lines can be
undetectable because of the low \met\ of the star, because the too-high (or
too-low) \temp\ does not populate enough of its electronic levels, 
or because the spectrum has a too-low STN.
This is illustrated in Fig~\ref{distr_stn_Nspec} and
Fig~\ref{distr_met_Nspec}, where the number of spectra having [X/H]
estimation diminishes with STN and \met . In general, only metal-rich stars
have elemental abundance estimations at any STN whereas metal poor stars have
estimations only if their spectra have high STN.

\subsection{Accuracy and reliability element by element}

We now discuss and summarize the reliability of the chemical
abundances for individual elements in light of the previous
discussion. First, we make some general remarks about the measured
elemental abundances and their errors:
\begin{itemize}
\item The accuracy of the chemical abundances depends on several
  variables. In particular if \temp\ is erroneously estimated, the
  abundances are affected to different degree for different
  elements. If there are systematics deviations in \temp\ there will be
  systematic deviations in [X/H] as well.
\item In general [X/H] are underestimated, and the magnitude of the
  bias is a function of \temp: stars with \temp\ $<$5000K yield on
  average abundances that are $\sim$0.1 dex lower than stars with
  $>$5000K, whose abundance errors average to zero. This effect
  increases the errors when stars with different \temp\ are
  analyzed. On the other hand, {\em relative abundances [X/Fe] are
    nearly unaffected because the trend is similar for all elements}.
\item The errors given below refer to the expected errors for
  two intervals of STN at \met\ $\sim$0.0 dex. As already discussed in
  the previous section, errors can increase for lower STN and lower
  \met.
\end{itemize}

\noindent {\bf Magnesium} yields reliable results on synthetic and
real spectra.  At any \temp\ we expect an abundance error $\sigma_{\rm
  Mg}\leq$0.15 dex for STN$\geq$40
and $\sim$0.25 dex for 20$\geq$STN$\geq$40.

\noindent {\bf Aluminum} abundances are obtained from only two
physically isolated lines (which are instrumentally blended at RAVE resolution).
Despite the instrumental blending, the lines are strong and give accurate estimates in
tests with synthetic and real spectra that shows no systematic offset.
For STN$\geq$40 we expect abundance errors $\sigma_{\rm Al}\leq$0.2
dex and $\sim$0.3 dex for 20$\geq$STN$\geq$40.

\noindent {\bf Silicon} is the most reliable element together with Fe.
We expect an abundance error $\sigma_{\rm Si}\leq$0.15 dex for
STN$\geq$40 and $\sim$0.20 dex for 20$\geq$STN$\geq$40.

\noindent {\bf Calcium} abundances are obtained from only five weak
\Cai lines.  \Cai is better measured at higher metallicity and \temp\
$<$5000K.  Estimated errors are $\sigma_{\rm Ca}\sim$0.25 dex for
STN$\geq$40 and $\sim$0.4 dex
for 20$\geq$STN$\geq$40.

\noindent {\bf Titanium} gives reliable abundances at high STN. At
STN$\sim$40 its abundance is reliable for \temp $<$5000K and
underestimated for higher temperatures. The correlation with \temp\
errors is particularly strong (\temp\ underestimation generates [Ti/H]
underestimation and vice versa), leading to larger errors. We expect
an abundance error of $\sigma_{\rm Ti}\sim$0.2 dex at STN$\geq$40 and
$\sim$0.3 dex for 20$\geq$STN$\geq$40.  We recommend separately analyzing 
stars with \temp\ lower and higher than 5000K.

\noindent {\bf Iron} is the most reliable element together with Si. It
can be accurately measured on spectra with any \temp\ in the range we
consider.  We expect an abundance error of $\sigma_{\rm Fe}\sim$0.1
dex at STN$\geq$40 and $\sim$0.2 dex at 20$\geq$STN$\geq$40.

\noindent {\bf Nickel} has six weak lines in the RAVE wavelength range
that are visible only at \temp$<$5000K.  In synthetic spectra with
STN=100 the pipeline yields errors comparable to those of other
elements but biased to lower values by $\sim$0.2 dex.  
It is not measurable for \met$<$-0.6 dex.  Tests on real
spectra are inconclusive because they were all performed for stars
with \temp $>$5000K. We expect uncertainties of $\sigma_{\rm
  Ni}\sim$0.2 dex for STN$\geq$40 and $\sim$0.3 dex for
20$\geq$STN$\geq$40.  Because of the small number and the weakness of its
lines, it is advisable to use Ni values with care.

\noindent The {\bf overall metallicity} \Mchem\ is inferred from the
chemical elemental abundances with Equation~\ref{eq_rave_met} \citep[Salaris
et al.][]{salaris}, with the abundances [Mg/H], [Si/H] and [Fe/H]
  delivered by the chemical pipeline. The metallicity \Mchem\ is
  reliable and we expect errors of $\sigma_{\rm [m/H]}\sim$0.1 dex for
  STN$\geq$40 and $\sim$0.2 dex for 20$\geq$STN$\geq$40.

  \noindent {\bf The $\alpha$ enhancement} [$\alpha$/Fe] has been
  computed as the average of [Mg/Fe] and [Si/Fe]
  (Equation~\ref{alpha_enh}). In the range 20$\geq$STN$\geq$40 it is 
  advisable to use it instead of the
  single $\alpha$ element values because it is more reliable.
  We estimate errors of
  $\sigma_{\alpha}\sim$0.1 dex for STN$\geq$40 and $\sim$0.2 dex for
  20$\geq$STN$\geq$40.

\subsection{The data}
\begin{figure}[t]
\centering
\includegraphics[width=12cm,clip]{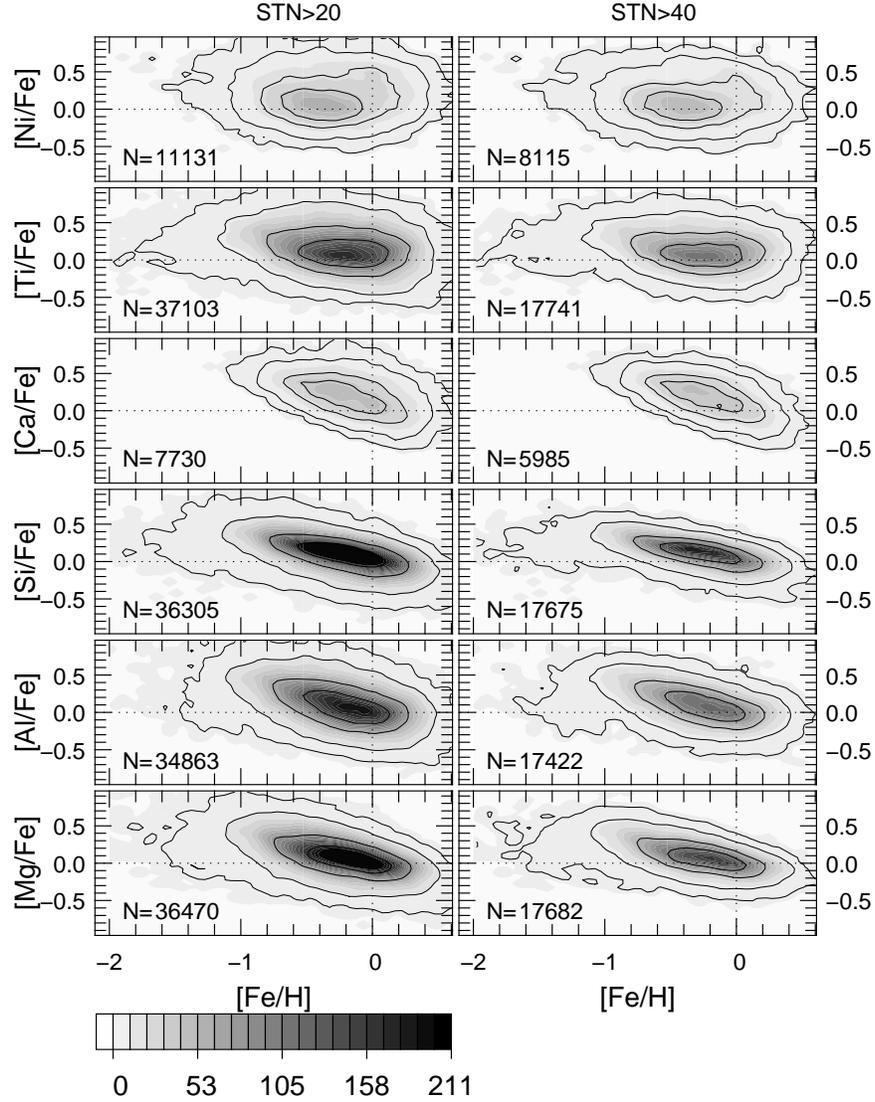}
\caption{Abundances relative to iron for spectra with STN$>$20 (left panels)
and STN$>$40 (right panels). The isocontours hold 34.0\%, 68.0\%, 95.0\% and
99.5\% of the sample.}
\label{chem_distr} 
\end{figure}

The RAVE chemical catalogue contains data for 37,848 spectra of 36,561 stars
and it is provided as an ASCII table of 37,848 lines.
It contains chemical abundances for the elements 
Mg, Al, Si, Ca, Ti, Fe and Ni, RAVE stellar parameters, signal-to-noise STN, 
object name and other parameters for quality checks as
explained in Table~\ref{tab_cat}. The distributions of the relative chemical
abundances are shown in Figure~\ref{chem_distr}.
The catalogue will be accessible online at \url{http://www.rave-survey.org} and
via the Strasbourg Astronomical Data Center (CDS) service.

\clearpage

\clearpage

\begin{table}
\caption[]{Catalogue description}
\label{tab_cat}
\vskip 0.3cm
\centering
\begin{tabular}{llcl}
\hline
\noalign{\smallskip}
Field number & name         & null value & description  \\
\noalign{\smallskip}
\hline
\noalign{\smallskip}
1 & Object ID & ... &  Identifier of the star\\
2 & [Mg/H] & -9.99 & Mg abundance\\
3 & N & ... & number of Mg lines measured\\
4 & [Al/H] & -9.99 & Al abundance\\
5 & N & ... & number of Al lines measured\\
6 & [Si/H] & -9.99 & Si abundance\\
7 & N & ... & number of Si lines measured\\
8 & [Ca/H] & -9.99 & Ca abundance\\
9 & N & ... & number of Ca lines measured\\
10 & [Ti/H] & -9.99 & Ti abundance\\
11 & N & ... & number of Ti lines measured\\
12 & [Fe/H] & -9.99 & Fe abundance\\
13 & N & ... & number of Fe lines measured\\
14 & [Ni/H] & -9.99 & Ni abundance\\
15 & N & ... & number of Ni lines measured\\
16 & \temp\ & ... & RAVE effective temperature\\
17 & \logg\ & ... & RAVE gravity\\
18 & \Mrave\ & ... & RAVE metallicity\\
19 & \Mchem\ & ... & metallicity from the chemical pipeline\\
20 & \Achem\ & ... & $\alpha$-enhancement from the chemical pipeline\\
21 & STN & ... & signal-to-noise ratio\\
22 & frac & ... & fraction of spectrum matching the template well\\
23 & Ntot & ... & total number of lines measured\\
24 & $\chi^2$ & ... & $\chi^2$ between observed and template spectra\\
\noalign{\smallskip}
\hline
\end{tabular}
\end{table}

\section{Conclusions}\label{concl_sec}

This first release of the RAVE chemical catalogue reports the chemical
abundances of seven elements (Mg, Al, Si, Ca, Ti, Fe, Ni) for 36,561
stars (from 37,848 spectra) selected from the DR3 data release. The
chemical processing pipeline developed for the creation of the present
catalogue assures homogeneous elemental abundance measurements in addition to
the RAVE pipeline that estimates the values of the stellar parameters and radial velocities.  

The uncertainties in the estimated elemental abundances depend mostly on the
signal-to-noise ratio (STN) of the spectra; on the values of the stellar parameters
of the stars; and on the element considered.  The expected errors in
abundance can be as small as 0.1 dex at STN$>$80 for the elements Fe
and Si and up to $\sim$0.3 at 20$\geq$STN$\geq$40 for Ni.  

On average we expect errors of $\sim$0.2 dex at STN$\geq$40 for most
of the elements. The relative abundances [X/Fe] are reliable with an
underestimation of $\sim-0.1$ dex for [X/Fe]$>$+0.3.

We plan to increase the number of measured elements
in future data releases. We believe that with further development of
the chemical pipeline it should be possible to enrich the catalogue
with the abundances of at least two other elements (oxygen and
sulfur).

The availability of chemical elemental abundances, together with the radial
velocities of the stars (given in the RAVE DR3 data release) and their
distances (\citealp[Breddels et al.][]{breddels}, \citealp[Zwitter et
al.][]{zwitter2010}, \citealp[Burnett et al.][]{burnett}), make RAVE
a prime dataset for Galactic archaeology investigations.\\

C.B. would like to thank 
M. Asplund, D. Yong and F. Th\'evenin for the useful discussions he had with
them.

Funding for RAVE has been provided by: the Australian Astronomical
Observatory; the Lebniz-Institut für Astrophysik Potsdam (AIP); the
Australian National University; the Australian Research Council; the
European Research Council through ERC-StG 240271 (Galactica); the French
National Research Agency; the German Research Foundation (DFG); the
Sonderforschungsbereich ``The Milky Way System" (SFB 881) of the DFG;
the Istituto
Nazionale di Astrofisica at Padova; The Johns Hopkins University; the
National Science Foundation of the USA (AST-
0908326); the W.M. Keck foundation; the Macquarie University; the
Netherlands Research School for Astronomy; the Natural Sciences and
Engineering Research Council of Canada; the Slovenian
Research Agency; the Swiss National Science Foundation; the Science \&
Technology Facilities Council of the UK; Opticon; Strasbourg Observatory;
and the Universities of Groningen, Heidelberg
and Sydney.

The RAVE web site is at http://www.rave-survey.org. 

\newpage

\appendix
\section{Correction of the atomic oscillator strengths
in the RAVE/Gaia wavelength region}\label{app_linelist}

In the following we outline how the pipeline treats oscillator
strengths ($gf$, often expressed as a logarithm $\log gf$) in the
wavelength range 8400-8800\AA. These are necessary in order to obtain
trustworthy elemental abundance measurements in a wavelength region which is
still poorly known. According to the Vienna Atomic Line Database VALD
\citep[][]{vald} only 11 Fe lines have accurate $gf$-value laboratory
measurements (\citealp[Blackwell et al.,][]{blackwell_Fe},
\citealp[Bard et al.,][]{bard1}, \citealp[Bard \& Kock,][]{bard2}),
while the remaining hundreds of lines rely on theoretical and
semi-empirical calculations \citep[Kurucz,][]{kurucz}.  Some of these
values have proved imprecise, with errors as large as 1 dex for
important lines belonging to Fe and Si \citep[Bigot \&
Th\'evenin,][]{bigot}.  We have attempted to improve the oscillator
strength values of 604 absorption lines using an inverse spectral
analysis applied to spectra of the Sun and Arcturus.

\subsection{Method}

The RAVE line list consists of 604 absorption lines (excluding the \Hi
and \Caii strong lines) identified by looking at the high S/N Sun and
Arcturus spectra (\citealp[Hinkle et al.,][]{hinkle}) and verified by
cross checking the line identification of Moore et
al. \citep[][]{moore} and Hinkle et al.  \citep[][]{hinkle}. Some
previously unidentified lines have been added in order to match
features that are clearly visible in spectra of the Sun and Arcturus.

The lines have been selected from Kurucz data
\citep[Kurucz,][]{kurucz} and we have also adopted Kurucz's
wavelengths and excitation potentials $\chi$. Following Chen et
al. \citep[][]{chen} we also adopted the enhancement factor E$_\gamma$
for the Uns\"{o}ld approximation because it is commonly accepted that
the line broadening due to Van der Waals interaction obtained from
Uns\"{o}ld's approximation is too weak and needs to be corrected. For
the CN  molecule we used the dissociation energy D=7.75eV, as suggested
by Sauval \& Grevesse \citep{sauval}. 

We performed the inverse spectral analysis by synthesizing solar and
Arcturus's spectra with MOOG (a standard LTE line analysis and synthesis
code by \citealp[Sneden,][]{sneden}) and changing the $\log gf$s
values until a good match with both observed spectra is reached. We
show here that matching these two spectra (or more) allows us to
improve the $\log gf$s of blended lines, which otherwise would be
impossible to disentangle.

The EW of a measured absorption line is related to the $\log gf$ value
through the curve of growth equation \citep[see Gray,][formula
16.4]{gray_d}
\begin{equation}\label{for_grow}
\log \frac{EW}{\lambda}=\log C+\log A+ \log (gf\cdot\lambda)-\theta\chi-\log
\kappa_{\nu},
\end{equation}
where $C$ and $\theta$ are functions\footnote{ $C$ and $\theta$ are
expressed as follows
\begin{flushleft}
$C  =  constant\cdot \frac{\pi e^2}{mc^2}\frac{N_j/N_E}{u(T)} N_H$\\
$\theta  =  5040/T$\\
\end{flushleft}
\noindent
where T is  the temperature, $N_j/N_E$ is the fraction of  atoms of the j-th
stage  of ionization  with respect  to the  number of  atoms of  the element
considered, $N_H$ is the number of hydrogen atoms per unit volume and $u(T)$
is the partition function.}  of  the temperature T, and $\kappa_{\nu}$ is
the opacity at the  frequency $\nu$.
When all the atmosphere model parameters are fixed, the curve of
growth equation can be written as a function, $f$, of $\log gf$ and
elemental abundance $A$
\begin{equation}
EW=f(\log gf, A).
\end{equation}

This allows us to disentangle two or more blended lines: let
$EW_{blend}^{x}$ be the EW of a blended feature of the star $x$
composed of $n$ lines of $m$ different elements ($n\geq m$) and having
$EW_1, EW_2...EW_n$ unknown EWs.  Then we can write the system of
equations
\begin{eqnarray}\label{eq_syst}
EW_{blend}^{1} & = & \sum^{n} f(\log
gf_{n},A_{n})\\\nonumber
EW_{blend}^{2} & = & \sum^{n} f(\log gf_{n},A_{n})\\\nonumber
\vdots                &  & \vdots                       \\\nonumber
EW_{blend}^{N} & = & \sum^{n} f(\log gf_{n},A_{n})\\\nonumber
\end{eqnarray}
\noindent where $A_{n}$ represents the abundance of the element 
the line belongs to. 

Since the physical conditions and elemental abundance $A$ of each stars are in
general different, the values $EW_{blend}^{x}$ are different as well.
The equation system \ref{eq_syst} is solvable when the ($N$) number of
equations is greater than the unknown $n+m$.  The existence and
uniqueness of the solution is assured even for lines where the
weak-line approximation does not hold, because the first derivative of
the curve of growth is always positive.

This means that it is always possible to determine the $\log gf$s of a
blended feature and the elemental abundances by using a large enough number of
spectra of different stars. In our case we have used two spectra: the
Sun and Arcturus.

\subsection{The correction routine}

The following routine is valid when the stellar parameters and
elemental abundances of both stars are known. We also consider later the case where
elemental abundances are not known. To synthesize the solar spectrum we assumed an
effective temperature \temp=5777K, gravity \logg=4.44, microturbulence
V$_t$=1.0 km/s and metallicity \met=0.00.  For Arcturus the
atmospheric parameters are taken from Earle, Luck \& Heiter
\citep[][]{luck2005} (\temp=4340K, \logg=1.93, V$_t$=1.87 km/s
\met=--0.55 dex).  

The synthetic spectra have been obtained assuming solar chemical
abundances by Grevesse \& Sauval \citep[][]{grevesse} (hereafter GS98)
and adopting the grid ATLAS9 model atmospheres of Castelli \& Kurucz
\citep[][]{castelli}.  The $\log gf$s correction is performed using
the following procedure:
\begin{enumerate}
\item Synthesize the solar and Arcturus spectra from our line list.
\item Select the first line of the list
\item Integrate the
  absorbed flux of the observed and synthesized spectra over an
  interval spanning 0.4\AA\ around the center of the line (roughly 1
  FWHM, which takes into account most of the absorbed flux) and
  calculate the normalized EW residual (NEWR), defined as follows
\begin{equation}
NEWR=\frac{(F_{synt}-F_{obs})}{F_{obs}}
\end{equation}
where $F_{synt}$ and $F_{obs}$ are the flux of the synthetic and the
observed spectrum, respectively. For the Sun and Arcturus we write ($NEWR_{Sun}$) and
($NEWR_{Arc}$).
\item If $NEWR_{Sun}<-0.05$ then add $+0.01$ to the $\log gf$, else if
  $NEWR_{Sun}>0.05$ then add $-0.01$ to the $\log gf$ (i.e. match the
  Sun first). Go to step 7.
\item If $|NEWR_{Sun}|<0.05$ and $NEWR_{Arc}<-0.05$
then add $-0.01$ to the $\log gf$, else if
$|NEWR_{Sun}|<0.05$ and $NEWR_{Arc}>0.05$ then add
$+0.01$ to the $\log gf$ (e.g. if the Sun matches
 then try to match Arcturus). Go to step 7.
\item If $|NEWR_{Sun}|$ and $|NEWR_{Arc}|$ are both $<0.05$
then a valid $\log gf$ has been reached.
\item Select the next line from the line list and return to step 3.
\end{enumerate}

The routine is iterated until no more improvement is seen. For about
ten lines a good match could not be reached. Several reasons may be
behind this: bad continuum correction, lines located over the \Caii
lines where the fit is unaccurate, misidentified lines or unrecognized
blends. For these lines, we did a visual check and manually corrected
the $\log gf$ values.  The correction is good enough to derive
$gf$-values that yield abundance errors per measured line smaller than
0.2 dex in average for both the Sun and Arcturus.

Since the abundance of Arcturus is not known for all elements, we decided
to assume an initial abundance for all the elements ([X/H]=\met) and
find the right abundances with the following procedure:
\begin{enumerate}
\item Synthesize the Sun's and Arcturus's spectra.
\item Correct the $\log gf$ values by running the above correction routine
until no more improvements are possible.
\item Check the distribution of the NEWRs for each element
separately: if a positive/negative offset is present,
decrease/increase its abundance.
\item Synthesize the solar and Arcturus's spectra with the new $\log
gf$s and abundances.
\item Return to step 2.
\end{enumerate}
This algorithm is repeated until there are no more improvements.  The
results of this procedure are the new $\log gf$s and Arcturus's
chemical abundances (listed in Table~\ref{abundances}).

\subsubsection{Multiplets treatment}\label{sec_multiplets}

In the RAVE/Gaia  wavelength range there are features  that belong to atomic
multiplets. An  interesting characteristic of these multiplets  is that they
share the  same excitation  potential $\chi$ and  have small  differences in
wavelength.  In this  case, obtaining a proper $\log  gf$ for the individual
lines is problematic.  

We work around this problem in the following way: consider the two
\Ali lines at $\lambda_1=8773.896\AA$ and $\lambda_2=8773.897\AA$ that
share the same excitation potential $\chi=4.021947 eV$.  The two lines
have unknown equivalent widths $EW_1$ and $EW_2$ but the sum
\begin{equation}\label{ew_sum}
EW_{Tot}=EW_1+EW_2,
\end{equation}
is measured and well known. We can write the curve of growth as
\begin{equation}\label{eq_cog}
\frac{EW}{\lambda}=\frac{C\cdot A_{Al} \cdot (gf)\cdot \lambda}{10^{\theta \chi} \cdot \kappa_{\nu}},
\end{equation}

\noindent
where $A_{Al}$ is the aluminum abundance and the other quantities are
as in Eq.~\ref{for_grow}.  Substituting Eq.~\ref{eq_cog} in
Eq.~\ref{ew_sum} and adopting the approximation $\lambda_1 \simeq
\lambda_2 = \lambda$ we obtain
\begin{equation}
\frac{EW_{Tot}}{\lambda}=\frac{EW_1+EW_2}{\lambda}=\frac{C \cdot A_{Al} \cdot (gf_1+gf_2) \cdot \lambda}{10^{\theta \chi}
\cdot \kappa_{\nu}}.
\end{equation}
\noindent
Using the logarithmic form, this equation then becomes
\begin{equation}
\log \frac{EW_{Tot}}{\lambda}=\log C+\log A+ \log ((gf_1+gf_2) \cdot \lambda)-\theta\chi-\log \kappa_{\nu},
\end{equation}
\noindent
which  suggests that an appropriate functional form for a ``dummy" variable
$\log gf_{d}$ would be $\log (gf_1+gf_2)$.  
Even  if this does not  allow us to disentangle  the two $\log
gf$s, we can consider the multiplet as if it were one single line and correct
the $\log gf_{fic}$; it  will be a ``dummy'' value but it results in giving
the correct abundance of Al.  This correction has also been applied
to  three  \Mgi  multiplets,   centered  at  wavelengths  8717.814,  8736.020
and 8773.896$\AA$.

\subsection{Tests}

\begin{figure}[b]
   \centering 
   \includegraphics[bb=59 339 301 462,width=7cm,clip]{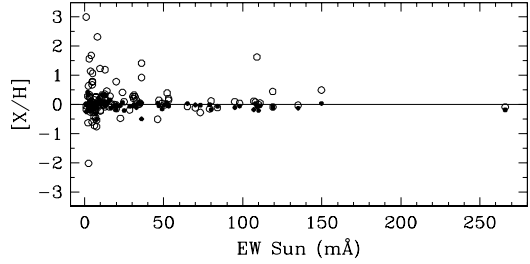}
      \caption{solar elemental abundances obtained from 100 atomic lines for which EWs (x-axis)
	have been measured by hand from the solar spectrum. The open points 
	represent the elemental abundances obtained by using the Kurucz $gf$-values,
	whereas the corrected $gf$-values are represented by the solid points.
	The respective dispersions are $\sigma$=0.63 dex (open points) and
	$\sigma$=0.15 dex (solid points).
              }
         \label{ew_sun} 
	\end{figure}

In order to test the accuracy of this procedure, we have measured by hand the EWs
of 100 atomic lines in the solar spectrum. Figure~\ref{ew_sun} presents
the difference in abundances obtained by using the Kurucz $gf$-values
(open points) and the corrected ones (solid points).  Our correction
reduced the standard deviations of the abundance residuals from
$\sigma$=0.63 dex to $\sigma$=0.15 dex and their averages from +0.16
to -0.05.  

\begin{figure}[t]
\centering 
\includegraphics[width=7cm,clip]{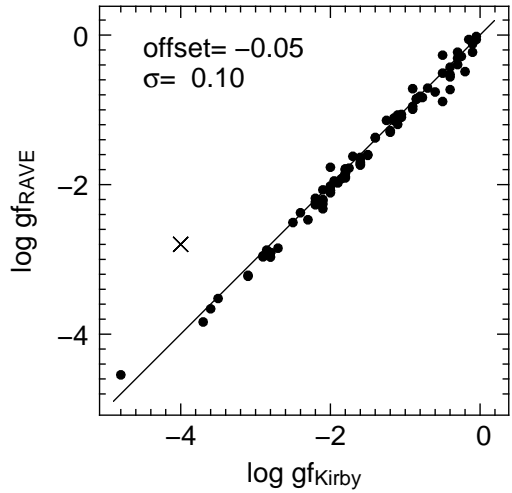}
\caption{
Comparison between corrected $\log gf$s by Kirby et al. 2008
(x-axis) and  RAVE (y-axis).  The statistics on the panel refers to
points after the crossed outlier has been removed.}
\label{loggf_kirby_us} 
\end{figure}

We also compared our results to the $\log gf$ by Kirby et
al. \citep[][]{kirby}, who corrected the $gf$-values by applying a
similar method, i.e., by synthesizing the solar and Arcturus's spectra and
comparing the synthetic spectra with the observed ones.  A comparison
between our and their results is shown in Figure~\ref{loggf_kirby_us}
for the 89 corrected lines in common: on average our $\log gf$ are
lower by --0.05 dex. This is very likely due to different solar elemental abundances
adopted: Kirby adopts the Anders \& Grevesse \citep[][]{anders}
abundances which are on average higher than Grevesse \& Sauval
\citep[][]{grevesse}, which we adopt.   

One more test involves comparing our results for Arcturus's elemental abundances with 
abundances quoted in the literature. This is done in Tab.  \ref{abundances}
where our elemental abundances are compared with those of Luck
\& Heiter \citep[][]{luck2005}, Th\'evenin \citep[][]{thevenin2}
(T98), and Fulbright et al. \citep[][]{fulbright} (F07).  From the
consistency of these tests, we are confident that our computed
$\log gf$ values are accurate and that our procedure yields robust and reliable
elemental abundance estimates.

\begin{table}[t]

\centering
\begin{tabular}{lcccc}
\hline
\noalign{\smallskip}
abundance & this work         & LH05  & T98 & F07\\
\noalign{\smallskip}
\hline
\noalign{\smallskip}
\mbox{[O/H]}    & +0.17      &                     &  --0.30  &--0.10\\
\mbox{[Mg/H]}   & --0.03      &  +0.02              &  --0.45 &--0.10\\
\mbox{[Al/H]}   & --0.27      &  --0.28              &  --0.40&--0.17\\
\mbox{[Si/H]}   & --0.15      &  --0.14              &  --0.30&--0.15\\
\mbox{[S/H] }   & --0.13      &                     &         &      \\
\mbox{[Ca/H]}   & --0.26      &  --0.56              &  --0.20&--0.28\\
\mbox{[Ti/H]}   & --0.22      &  --0.39              &  --0.20&--0.20\\
\mbox{[Cr/H]}   & --0.37      &  --0.55              &  --0.20&      \\
\mbox{[Fe/H]}   & --0.52      &  --0.55              &  --0.40&--0.54\\
\mbox{[Co/H]}   & --0.17      &  --0.36              &  --0.20&      \\
\mbox{[Ni/H]}   & --0.45      &  --0.48              &  --0.35&      \\
\mbox{[Zr/H]}   & --1.00      &                     &          &      \\
\noalign{\smallskip}
\hline
\end{tabular}
\caption[]{Arcturus's elemental abundances
compared with the results by 
Luck \& Heiter \citep[][]{luck2005} (LH05), Th\'evenin \citep[][]{thevenin2}
(T98), Fulbright et al. \citep[][]{fulbright} (F07)}
\label{abundances}
\end{table}

\begin{deluxetable}{crcrcc}
\tabletypesize{\scriptsize}
\tablecaption{Seven lines extracted from the RAVE line list. The whole table
contains 604 lines with $\log
gf$ correction, and the 3 \Caii and 8 \Hi lines which are not used for
elemental abundance measurements (and did not undergo to $\log gf$
corrections). The atomic species 607.0 represent the CN molecule, following
the MOOG format. \label{table_loggf}}
\tablewidth{0pt}
\tablehead{
\colhead{wavelength(\AA )} & \colhead{atomic species} & \colhead{$\chi$(eV)} &
\colhead{$\log gf$} & \colhead{E$_{\gamma}$} &\colhead{Dissociation energy
(eV)} 
}
\startdata
8643.306&  26.0&  4.913304& -2.470&  1.40& 0.00\\
8644.077& 607.0&  0.933782& -1.773&  2.50& 7.75\\
8646.050& 607.0&  1.324726& -1.540&  2.50& 7.75\\
8646.371&  14.0&  6.206423& -1.740&  1.30& 0.00\\
8647.175& 607.0&  0.877831& -1.872&  2.50& 7.75\\
8648.465&  14.0&  6.206423&  0.100&  1.30& 0.00\\
8648.688& 607.0&  0.910915& -1.516&  2.50& 7.75\\
\enddata
\tablecomments{Table~\ref{table_loggf} is published in its entirety in the
electronic edition of the {\it Astronomical Journal}.
}
\end{deluxetable}

\clearpage

\section{Microturbulence}\label{app_microt}
The polynomial function used to determine the microturbulence $\xi$
is expressed as follows
\begin{equation}\label{microt_for}
\xi_{poly}(\mbox{km s$^{-1}$})=\sum_{i,j=0,1,2,3}^{i+j \leq 3} a_{ij} T^i
(\log g)^j
\end{equation}
where the coefficients $a_{ij}$ are
\begin{eqnarray}
a_{00}&=&-10.3533\\\nonumber
a_{01}&=&2.59492\\\nonumber
a_{02}&=&0.161863\\\nonumber
a_{03}&=&0.176579  \\\nonumber
a_{10}&=&0.00509193  \\\nonumber
a_{11}&=&-0.00157151  \\\nonumber
a_{12}&=&-3.99782\cdot 10^{-7} \\\nonumber
a_{20}&=&-0.000361822  \\\nonumber
a_{21}&=&3.82802\cdot 10^{-7}  \\\nonumber
a_{30}&=&-5.18845\cdot 10^{-11} \\\nonumber
\end{eqnarray}

\clearpage


\begin{thebibliography}{}
   \bibitem[2004]{allende} Allende Prieto, C., Barklem, P.S.,
	Lambert, D.L., Cunha, K., 2004, A\&A 420, 183
   \bibitem[1989]{anders} Anders, E., Grevesse, N., 1989, Geochim. Cosmochim. Acta
	53, 197
   \bibitem[2005]{asplund_ab} Asplund, M., Grevesse, N., 
	Sauval, A. J., 2005, in ASP Conf.Ser. 336, Cosmic Abundances as
Record of Stellar Evolution and Nucleosynthesis, ed. Thomas G. Barnes III
and Frank N. Bash, (San Francisco: ASP), 25 
   \bibitem[2009]{abazajian}  Abazajian, K.N., et al.,  2009, ApJS 182, 543
   \bibitem[1991]{bard1} Bard, A., Kock, A., Kock, M. 1991,
	 A\&A, 248, 315
   \bibitem[1994]{bard2} Bard, A., Kock, M. 1994,
	 A\&A, 282 ,1014
   \bibitem[2000]{barklem} Barklem, P.S., Piskunov, N., O'Mara, B.J.,
    2000, A\&AS 142, 467
   \bibitem[2005]{bensby2005} Bensby, T., Feltzing, S., Lundstr\"om, I.,
	Ilyin, I., 2005, A\&A 433, 185
   \bibitem[2006]{bigot} Bigot, L., Th\'evenin, F. 2006,
	 MNRAS 372, 609
   \bibitem[1986]{blackwell_Fe} Blackwell, D. E., Booth, A. J.,
	Haddock, D. J., Petford, A. D., Leggett, S. K. 1986,
	MNRAS, 220, 549
   \bibitem[2010]{Boeche_PhD} Boeche, C., PhD thesis, 2011,
	urn:nbn:de:kobv:517-opus-52478, Potsdam
	Universit\"at, \url{http://opus.kobv.de/ubp/volltexte/2011/5247/}
   \bibitem[2003]{borrero} Borrero, J. M., Bellot Rubio, L. R., Barklem,
	P. S., del Toro Iniesta, J. C., 2003
	A\&A 404, 749
   \bibitem[2010]{breddels} Breddels, M. A., et al., 2010, A\&A 511, 90 
   \bibitem[2011]{burnett} Burnett B., et al.,  2011, A\&A 532, 113 
   \bibitem[2003]{castelli}   Castelli, F., Kurucz, R. L.,
	2003, in IAUS symp. 210, Modelling of Stellar Atmospheres, ed. N.
Piskunov, W.W. Wiess and D.F. Gray , Published on behalf of the IAU by the
ASP, 20
  \bibitem[2000]{chen} Chen, Y.Q., Nissen, P.E., Zhao, G., Zhang, H.W.,
	Benoni, T., 2000, A\&AS 141, 491
   \bibitem[1993]{edvardsson} Edvardsson, B., Andersen, J., Gustafsson, B.,
	Lambert, D.L., Nissen, P.E., Tomkin, J., 1993, A\&A 275, 101
   \bibitem[1965]{eggen} Eggen, O., 1965, in Galactic Structure, Stars and Stellar
Systems, Volume 5, ed. A. Blaauw \& M. Schmidt, (Chicago: Univ. of Chicago
Press), 111 
   \bibitem[1962]{ELS} Eggen, O., J., Lynden-Bell, D., Sandage, R., 1962, ApJ
	136, 748
   \bibitem[2002]{freeman} Freeman, K., Bland-Hawthorn, J., 2002, ARA\&A 40,
	487
   \bibitem[2008]{freeman1} Freeman, K., Bland-Hawthorn, J., 2008, in ASP
Conf. Ser. 399, Panoramic Views of Galaxy Formation and Evolution, ed. 
Tadayuki Kodama, Toru Yamada, and Kentaro Aoki, (San Francisco: ASP), p.439
\bibitem[1998] {fuhrmann1998}Fuhrmann, K., 1998, A\&A 338, 161
\bibitem[2008] {fuhrmann2008} Fuhrmann, K., 2008, MNRAS 384, 173 
   \bibitem[2007]{fulbright} Fulbright, J.P., McWilliam, A.,
        Rich, R.M., 2007, AJ 661, 1152
   \bibitem[2006]{fulbright2006} Fulbright, J.P., McWilliam, A.,
	Rich, R.M., 2006, AJ 636, 821
  \bibitem[2005]{gray_d} Gray, D. F., 2005,
	{\it The Observation and Analysis of Stellar
	Photospheres}, Third Edition, Cambridge University
	Press, New York
   \bibitem[1983]{gr} Gilmore, G., Reid, N., 1983, MNRAS 202, 1025
   \bibitem[1989]{gwk} Gilmore, G., Wyse, R.F.G., Kuijken, K.,
 1989, ARA\&A 27, 555 
   \bibitem[1998]{grevesse} Grevesse, N., Sauval, A.J. 1998,
	Space Sci. Rev. 85, 161
   \bibitem[1981]{gurtovenko} Gurtovenko, E.A., Kostik, R. I. 1981,
	A\&AS, 46, 239
   \bibitem[2006]{helmi} Helmi, A., Navarro, J.F., Nordstr\"om, B.,
Holmberg, J., Abadi, M.G., Steinmetz, M., 2006, MNRAS 365, 1309
   \bibitem[2000]{hinkle}   Hinkle, K., Wallace, L.,
	Valenti, J.,; Harmer, D., 2000,
	{\it Visible and Near Infrared Atlas of the Arcturus
	Spectrum 3727-9300 $\AA$}, Available at:
	ftp://ftp.noao.edu/catalogs/arcturusatlas/visual
\bibitem[2000]{hog} H\o g, E.; Fabricius, C.; Makarov, V. V.; Urban, S.;
	Corbin, T.; Wycoff, G.; Bastian, U.; Schwekendiek, P.; Wicenec, A. 2000,
	A\&A 355, L27
   \bibitem[2008]{kirby} Kirby, E.N., et al., 2008, AJ 682, 1217 
   \bibitem[1995]{kurucz} Kurucz, R. L. 1995, in ASP Conf.Ser 78,
   Astrophysical Application of Powerful New Database, ed. S.J. Adelman,
   W.L. Wiese, (San Francisco, CA), 205
   \bibitem[1999]{vald} Kupka, F.,Piskunov, N., Ryabchikova, T.A.,
Stempels, H.C., Weiss, W.W., 1999, A\&AS, 138, 119
   \bibitem[2005]{luck2005} Luck, R.E., Heiter, U. 2005,
	2005, AJ 129, 1063
\bibitem[2006]{luck2006} Luck, R.E., Heiter, U., 2006, AJ 131, 3069
\bibitem[2007]{luck2007} Luck, R.E., Heiter, U., 2007, AJ 133, 2464
   \bibitem[2003]{mishenina} Mishenina, T.V., Kovtyukh, V.V.,
     Korotin, S.A., Soubiran, C., 2003, Astron. Zh. 80, 458
   \bibitem[1966]{moore} Moore, C. E., Minnaert, M. G. J., 
	Houtgast, J., 1966, {\it The solar spectrum 2935 $\AA$ to 8770
	$\AA$}, National Bureau of Standards Monograph 61
   \bibitem[2004]{nordstrom} Nordstr\"om, B., et al., 2004, A\&A 418, 989
   \bibitem[2001]{gaia} Perryman, M. A. C., et al., 2001,
    A\&A 369, 339
   \bibitem[2003]{reddy2003} Reddy B.,E., Tomkin J., Lambert L., and
	Allende Prieto C., 2003, MNRAS 340, 304 
   \bibitem[2006]{reddy} Reddy, B.E., Lambert D.L., Allende Prieto C., 2006,
    MNRAS 367, 1329
   \bibitem[2003]{robin} Robin, A., C., Reyl\'e, C., Derri\`ere, S., Picaud, S.,
	2003, A\&A 409, 523
   \bibitem[2008]{roeser} Roeser, S., Schilbach, E., Schwan, H., Kharchenko,
	N.V.,  Piskunov, A.E., Scholz R.-D., 2008, A\&A 488, 401
   \bibitem[2010]{ruchti} Ruchti, G. R. et al., 2010, ApJ 721, 92
  \bibitem[1993]{salaris} Salaris, M., Chieffi, A., Straniero, O., 1993, 
	AJ 414, 580
   \bibitem[1994]{sauval} Sauval, A. J., Grevesse, N.,
	1994, in IAUS symp. 154, Infrared solar physic, ed. D.M. Rabin, John
T. Jefferies, and C. Lindsey, (Kluwer Academic Publishers; Dordrecht), 549
   \bibitem[2011]{siebert} Siebert, A., et al., 2011, AJ 141, 187
   \bibitem[1973]{sneden} Sneden, C., 1973, Ph.D. thesis, Univ. Texas ar
	Austin, software available at
\url{http://verdi.as.utexas.edu/moog.html}
\bibitem[2005]{soubiran} Soubiran, C., Girard, P., 2005, A\&A 438, 139
\bibitem[2010]{soubiran2010} Soubiran C., Le Campion J.-F., Cayrel de
Strobel G., and Caillo A., A\&A submitted, arXiv: 1004.1069
   \bibitem[2006]{steinmetz} Steinmetz, M., et al., 2006,
   AJ 132, 1645
   \bibitem[1989]{thevenin} Th\'evenin, F. 1989, 
	A\&AS, 77, 137
   \bibitem[1990]{thevenin1} Th\'evenin, F. 1990, 
	A\&AS, 82, 179
   \bibitem[1998]{thevenin2} Th\'evenin, F. 1998,
	Bull. Inf. CDS 49 (unpublished)
   \bibitem[1955]{unsold} Uns\"{o}ld, A., 1955, Physik der Sternatmospharen. Springer, Berlin
\bibitem[2005]{valenti} Valenti, J.A., Fisher, D.A., 2005, AJSS 159, 141
\bibitem[2004]{venn}  Venn K.A., Irwin M., Shetrone M.D., Tout C.A., Hill
V., Tolstoy E., AJ 128, 1177
   \bibitem[2009]{yanny}Yanny, B., et al., 2009, AJ 137, 4377
   \bibitem[2008]{zwitter} Zwitter, T., et al., 2008, AJ 136, 421
   \bibitem[2010]{zwitter2010} Zwitter, T., et al., 2010, A\&A 522, 54
   \bibitem[2004]{zacharias} Zacharias, N., Urban, S.E., Zacharias, M.I.,
   Wycoff, G.L., Hall, D.M., Germain, M.E., Holdenried, E.R., Winter, L.
   AJ 127, 3043
\end{thebibliography}
\end{document}